%% file: main.tex
\documentclass[sigconf, nonacm]{acmart}
\usepackage{tsexplain}

\setcopyright{none} 
\begin{document}
\title{\sys: Explaining Aggregated Time Series \\ by Surfacing Evolving Contributors}
\subtitle{Technical Report} 
\subtitlenote{Please see our published version in ICDE 2023.}

\author{Yiru Chen}
\email{yiru.chen@columbia.edu}
\affiliation{
  \institution{Columbia University}
  \city{New York}
  \state{NY}
  \country{USA}
}

\author{Silu Huang}
\email{silu.huang@microsoft.com}
\affiliation{
  \institution{Microsoft Research}
  \city{Redmond}
  \state{WA}
  \country{USA}
}

\begin{abstract}
    Aggregated time series can be generated effortlessly everywhere, e.g., ``total confirmed covid-19 cases since 2019'', ``S\&P500 during the year 2020'', and ``total liquor sales over time''. Understanding ``how'' and ``why'' these key performance indicators(KPI) evolve over time is critical to making data-informed decisions. Existing explanation engines focus on explaining the difference between two relations. However, this falls short of explaining KPI's continuous changes over time, as it overlooks explanations in between by only looking at the two endpoints.  Motivated by this, we propose \sys, a system that explains aggregated time series by surfacing the underlying {\em evolving} top contributors. Under the hood, we leverage the existing work on {\em two-relations diff} as a building block and formulate a {\em $K$-Segmentation} problem to segment the time series such that each segment after segmentation shares consistent {\em explanations}, i.e., contributors. To quantify consistency in each segment, we propose a novel within-segment variance design based on top explanations; to derive the optimal {\em $K$-Segmentation} scheme, we develop a dynamic programming algorithm.  Experiments on synthetic and real-world datasets show that our explanation-aware segmentation can effectively identify evolving explanations for aggregated time series and outperform explanation-agnostic segmentation. Further, we proposed an optimal selection strategy of $K$ and several optimizations to speed up \sys for interactive user experience, achieving up to $13\times$ efficiency improvement.
  
  \end{abstract}

\maketitle

\input{content/intro}
\input{content/relatedwork}
\input{content/probOverview}

\input{content/design}
\input{content/efficiency}

\input{content/extension}
\input{content/exp}

\input{content/discussion.tex}
\input{content/conclusion}

\newpage
\bibliographystyle{ACM-Reference-Format}
\bibliography{ref}

\end{document}

%% file: content/intro.tex
\section{Introduction}\label{sec:intro}

Time series data is gaining increasing popularity these days across sectors ranging from finance, retail, IoT to DevOps. Time series analysis is crucial for uncovering insights from time series data and helping business users make data-informed decisions. 
A business analyst typically focuses on three questions: ``what happened'' to understand the changes in key performance indicator (KPI), ``why happened'' to reason why KPI changes, and ``now what''~\cite{sisu-decision-intelligence} to guide what actions should be taken. 
``What'' questions have been extensively studied both academia-wise~\cite{gray1997data} and industry-wise~\cite{Tableau,Powerbi,googletrend}. ``Why'' questions are starting to attract wide attentions~\cite{Wu2013ScorpionEA, Wang2015DataXA,Bailis2017MacroBasePA}.
Existing explanation engines focus on explaining (1) one aggregated value~\cite{joglekar2017interactive, googletrend} or (2) differences between two given relations: a test relation and a control relation~\cite{sarawagi2001idiff, Wu2013ScorpionEA, Bailis2017MacroBasePA, roy2014formal, Wang2015DataXA, Abuzaid2018DIFFAR, Ruhl2018TheCA, miao2019lensxplain, li2021putting, Tableau19ExplainData, PowerbiKeyInfluencers, Sisu, Imply, googletrend}.

However, KPIs are typically monitored continuously, reporting some aggregated time series.
Simply explaining one aggregated value overlooks the trend of time series, e.g. ``why up/down”; merely focusing on its two endpoints and explaining their difference overlooks the evolving explanations in between.
 As evidence, although {\tt key influencer} feature which explains the difference between two given relations is well received in PowerBI community, a highly voted feature request in PowerBI Idea Forum is called {\it key influencer\footnote{Influencer corresponds to explanation in our notion.} over time}~\cite{KeyInfluencerOverTime}. Below are some quotes from the user: {\it ``the mix of factors/influencers tends to be more dynamic than static over time...It would be nice to add a time dimension to the Key Influencers analysis to understand how the top key influencers evolve over the evaluated period"~\cite{KeyInfluencerOverTime}.}
In time-series scenarios, it is often more desired to explain the evolving dynamics over time than only to consider two end timestamps.

\stitle{Disclaimer.}
We remark that identifying the root cause of ``why'' questions in general is only plausible when combining human interpretations with tools. Quoted from Tableau \code{ExplainData} homepage~\cite{Tableau19ExplainData}: ``The tool uncovers and describes relationships in your data. It can't tell you what is causing the relationships or how to interpret the data.'' Following the literature~\cite{Abuzaid2018DIFFAR, Wu2013ScorpionEA, Wang2015DataXA, Bailis2017MacroBasePA}, the \texttt{explanation} in our work does not equate to ``cause'', instead it corresponds to the data slice that contributes most to the overall change as we will describe in our Background Section (Definition~\ref{def:explanation}).

\begin{figure}[t!]
    \centering
    \begin{subfigure}[t]{0.24\textwidth}
        \centering
        \includegraphics[height=1.43in]{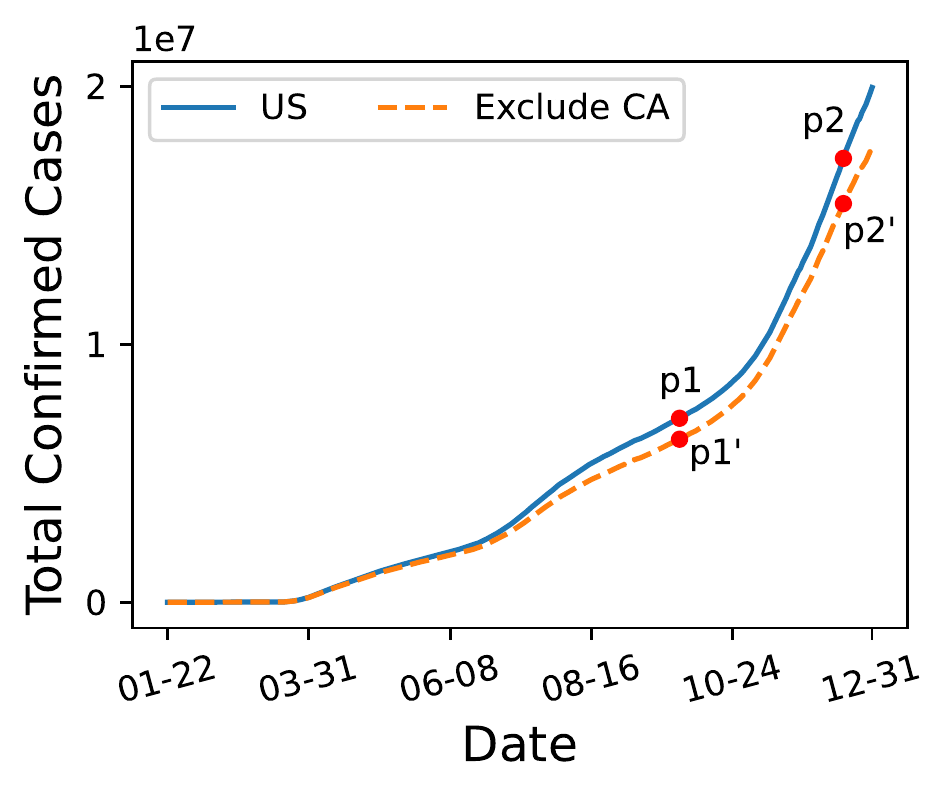}

        \caption{Total Confirmed Cases}
        \label{fig:total-covid-us}
    \end{subfigure}%
    ~
    \begin{subfigure}[t]{0.24\textwidth}
        \centering
        \includegraphics[height=1.43in]{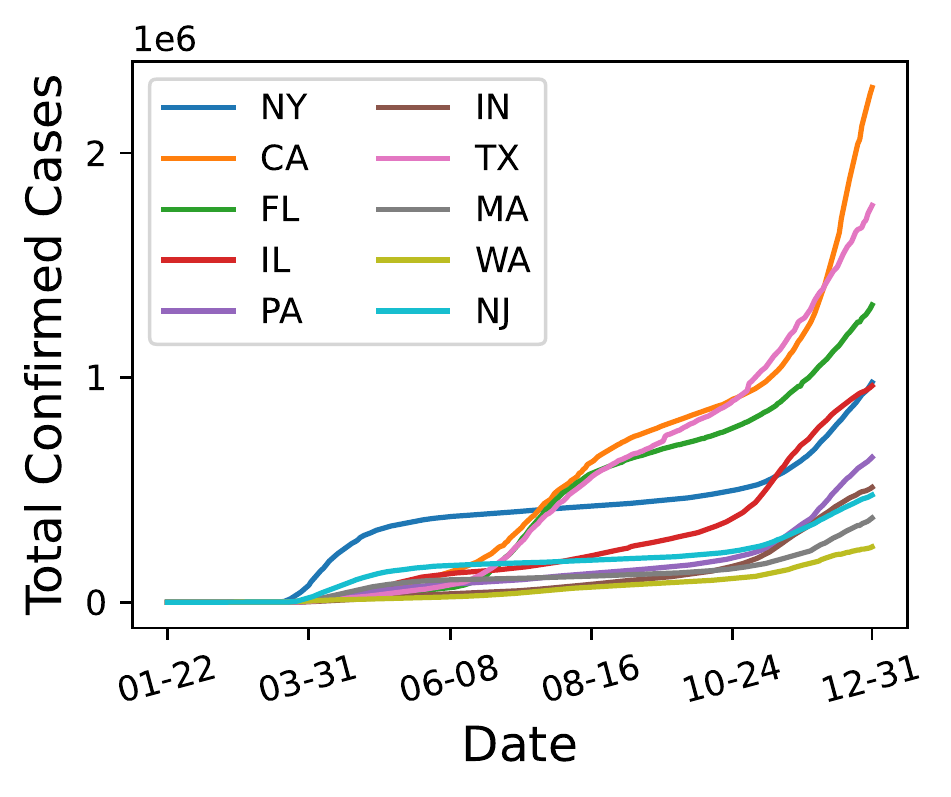}
        \caption{10 Example States}
        \label{fig:total-covid-states}
    \end{subfigure}
    \caption{COVID-19 Total Confirmed Cases~\cite{COVID-19-Johns-Hopkins}}
    \label{fig:total-covid}
\end{figure}

\stitle{Motivating Examples.}
We now describe three application scenarios of explaining changes in continuously evolving KPIs.

\bulletP{COVID-19.} Figure~\ref{fig:total-covid-us} depicts the total number of covid-19 confirmed cases during year 2020. This time series is obtained by performing a groupby-aggregate query over the original table~\cite{COVID-19-Johns-Hopkins}, which consists of attributes like {\small \texttt{state}, \texttt{total\_confirmed\_cases}, and \texttt{daily\_confirmed\_cases}}. By looking at Figure~\ref{fig:total-covid-us}, users can get an understanding of how overall trend evolves over time. One natural followup question is ``what makes the increase'' -- how different states contribute to the increase as time went along? Manual drill-down and browsing are laborious and overwhelming especially when there is a large number of attributes and each attribute is of high cardinality. Figure~\ref{fig:total-covid-states} illustrates a sample drill-down view along attribute {\small \texttt{state}}: first, looking at these sampled 10 states is already distracting, let alone full 58 states in the US; more importantly, it is still not clear what the answer to the above question is, though we do observe that each state contributes differently as time moves along. For instance, {\small \texttt{state=NY}} drives the initial outbreak in the US, while {\small \texttt{state=CA}} contributes most during the end of 2020.

\bulletP{S\&P500.} {\small \texttt{S\&P500}}~\cite{sp500chart} is a stock market index tracking the performance of around 500 companies in the US.
In a nutshell, {\small \texttt{S\&P500}} is calculated as the weighted average of these companies' stock prices.
After seeing how {\small \texttt{S\&P500}} evolves during 2020, users might be interested in explaining the movement of {\small \texttt{S\&P500}} by stock category. Intuitively, different stock categories drive the drop and rebound of {\small \texttt{S\&P500}}. For instance, \texttt{Category=Financial} plays a significant role in the drop during early covid-19 outbreak, but does not contribute to the rebound that much during the second quater of 2020.

\bulletP{Liquor-sales.}
The {\tt liquor-sales} time series corresponds to query {\small{\tt SELECT date, SUM(Bottles\_Sold) FROM Liquor GROUP BY date}}, where each row in relation {\small \texttt{Liquor}} represents a liquor purchase transaction with attributes including {\small \texttt{date}}, {\small \texttt{Bottle\_Volume(ml)}}, {\small \texttt{Pack}}, {\small \texttt{Category\_Name}}, and {\small \texttt{Bottles\_Sold}}.
Data analysts may wonder why the total bottles sold turns up since mid-January-2020 and how drinking behavior changes during pandemic. As we will show in our experiment, it turns out that people favors large pack liquor during pandemic, leading to sharp sales increase of {\small \texttt{Pack}=12} and {\small \texttt{Pack}=24}; and that people increase the purchase of large volume liquor such as {\small \texttt{Bottle\_Volume(ml)=750}} and {\small \texttt{Bottle\_Volume(ml)=1750}}.

\stitle{Problem and Challenges.}
These motivating examples can all be abstracted as the same problem: given a relation $R$, a set of explain-by attributes from $R$, and a time series aggregated from $R$, identify the top \texttt{explanations}, i.e., conjunctions of predicates over explain-by attributes, that contribute to the changes in the given aggregated time series.
Let us illustrate the motivating example of \texttt{covid-19} using this problem formulation.
Given a relation \texttt{Covid-19}, where each row records the total number of confirmed cases in a state on a particular date, a set of explain-by attributes, e.g., [{\small \texttt{state}}], and a time series aggregated from $R$, e.g., Figure~\ref{fig:total-covid-us} corresponding to query ``{\small{\tt SELECT date, SUM(total\_confirmed\_cases) FROM Covid-19 GROUP BY date}}'', our goal is to identify \texttt{explanations}, e.g., {\small \texttt{state=NY}}, that explains the surge in Figure~\ref{fig:total-covid-us}.
There are two main challenges in solving this problem: {\em (a)} \expls evolve over time; {\em (b)} interactivity is critical for data exploration and analytics.

\underline{Challenge (a)}: \expls tend to change over time. For instance, by looking at Figure~\ref{fig:total-covid-states}, we can observe that the increase in {\small \texttt{New York(NY)}} is the main reason of the total increase in Figure~\ref{fig:total-covid-us} during \texttt{2020-04} and \texttt{2020-05}, while {\small \texttt{California(CA)}} is the driving force during December 2020. Similarly, different stock categories are responsible for the surges and declines of \texttt{S\&P500} during different periods. Specifically, \texttt{technology} and \texttt{financial} industries play leading role in the sink of \texttt{S\&P500} during the initial covid-19 outbreak (around \texttt{2020-02-06} to \texttt{2020-03-24}); while \texttt{technology} industry is the top-contributor for \texttt{S\&P500}'s bounce back since \texttt{2020-03-24}, but \texttt{financial} industry is not. Having observed that \expls evolve along the time, our first technical challenge lies in how to identify time period with consistent \expls and how to derive \expls for each consistent time period.

\underline{Challenge (b)}: data analysts typically explore ``what'' and ``why'' questions iteratively to understand data and uncover insights. Studies~\cite{DBLP:journals/tvcg/LiuH14} have shown that low latency is critical in fostering user interaction, exploration, and the extraction of insights. It is desired that each query, including both ``what'' and ``why'' queries, can get answered in a second to ensure interactivity. Thus, how to reduce the latency for deriving explanations poses another challenge.

\stitle{Prior Works.}
``Why'' questions are gradually gaining attraction both academia-wise and industrial-wise. However, instead of explaining the continuous changes in a time series, existing works focus on explaining either (1) one aggregated value~\cite{joglekar2017interactive}, e.g., point $p_1$ in Figure~\ref{fig:total-covid-us}; or (2) the difference between two given relations~\cite{Wu2013ScorpionEA,Bailis2017MacroBasePA, li2021putting, sarawagi2001idiff, Tableau19ExplainData, PowerbiKeyInfluencers}, e.g., a test relation and a control relation corresponding to point $p_1$ and $p_2$ respectively in Figure~\ref{fig:total-covid-us}.
Reiterating the ``Disclaimer'' above, these tools can recommend and expedite answering ``why'' questions, but the human-in-the-loop interpretation is still required for true root cause analysis. To summarize, no prior works have studied the problem of explaining the evolving changes in time series as depicted in our motivating examples. Please refer to Section~\ref{sec:related} for detailed comparison.

\begin{figure}
    \centering
    \includegraphics[width=\linewidth]{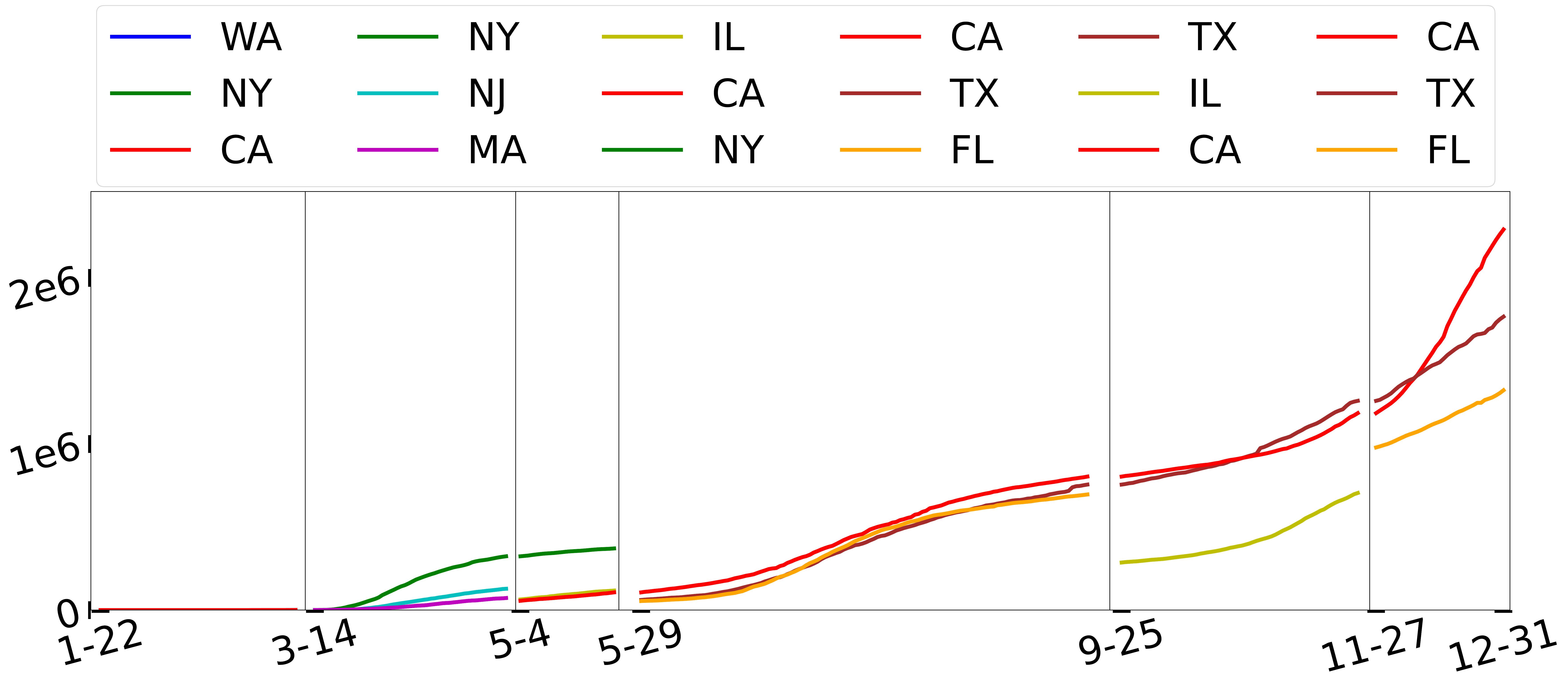}
    \caption{Evolving Explanations of Figure~\ref{fig:total-covid-us}.}
    \label{fig:covid-explain}
\end{figure}

\stitle{Our Solution.}
We propose \sys, a system to explain the continuous changes in aggregated time series. In \sys, given a relation, users can freely perform OLAP operations, including drill-down, roll-up, slicing, and dicing, and visualize what has happened to some KPI. To explain, users can then specify the time period they are interested as well as a set of explain-by attributes based on their domain knowledge. For the \texttt{COVID-19} example, \sys returns a trendline visualization in \Cref{fig:covid-explain}, where the whole input time series get partitioned into a few non-overlapping time periods and each time period is associated with the KPI trendlines for top \expls.
We can see that in early stage, NY and WA are the main contributors to the case increase; while CA, TX, IL,and FL become the main contributors later 2020.

Technically, to tackle challenge (a) of deriving {\em evolving explanations}, we formulate a {\em K-Segmentation} problem, aiming to partition the input time period into $K$ smaller time periods such that each time period shares consistent top \expls. Furthermore, since $K$ is hard to specify in practice, \sys employs ``Elbow method''~\cite{satopaa2011finding} for identifying the optimal $K$. We demonstrate the effectiveness of our problem formulation with both synthetic and real-world datasets. As for challenge (b) of interactivity, we first analyze the complexity of each step in \sys, identifying the bottleneck in the whole pipeline. Next, we propose several optimizations for reducing the bottlenecks in \sys. In our experiments, \sys has successfully answered all queries within one second.

\stitle{Contributions.} The contributions of this paper are as follows:
\begin{itemize}[leftmargin=*,noitemsep,topsep=0pt,parsep=0pt,partopsep=0pt]
    \item We formulate $K$-Segmentation problem for explaining the continuous changes in an aggregated time series. (Section~\ref{sec:prob})
    \item We propose a novel within-segment variance design based on top explanations to quantify consistency in each segment and experimentally prove its effectiveness. (Section~\ref{sec:design})
    \item We develop a dynamic program algorithm, perform complexity analysis, propose optimal selection strategy of $K$, and develop several optimizations for improving efficiency. (Section~\ref{sec:sys}) %
    \item We conduct experiments on both synthetic and real-world datasets, demonstrating the effectiveness and efficiency of \sys. (Section~\ref{ssec:exp-variance} and~\ref{sec:exp})
\end{itemize}

%% file: content/relatedwork.tex
\section{Related Work}\label{sec:related}
\vspace{-2mm}
\stitle{Data Explanation}
Existing data explanation engines mainly focuses on explaining (1) one aggregated value~\cite{joglekar2017interactive, googletrend}  or (2) the difference between two relations~\cite{sarawagi2001idiff, Wu2013ScorpionEA, Bailis2017MacroBasePA, roy2014formal, Wang2015DataXA, Abuzaid2018DIFFAR, Ruhl2018TheCA, miao2019lensxplain, li2021putting, Tableau19ExplainData, Sisu, Imply, PowerbiKeyInfluencers, googletrend}.
In academia, SmartDrillDown~\cite{joglekar2017interactive} explains one aggregated value by identifying explanations (called rules in SmartDrillDown) that have high aggregate value.
IDIFF~\cite{sarawagi2001idiff} identifies the differences between two instances of an OLAP cube.
Scorpion~\cite{Wu2013ScorpionEA} and Macrobase~\cite{Bailis2017MacroBasePA} aim to find the difference between outlier and inlier data.
RSExplain~\cite{roy2014formal} proposes an intervention-based framework to explain why SQL expression's result is high (or low). X-Ray~\cite{Wang2015DataXA} tries to reveal the common properties among all incorrect triples versus correct triples. Abuzaid et al.~\cite{Abuzaid2018DIFFAR}  unified various explanation engines and abstracted out a logical operator called diff. The cascading analysts algorithm~\cite{Ruhl2018TheCA} provides top {\em non-overlapping} explanations accounting for the major difference of two specified sets. Li et al.~\cite{li2021putting} also compares two set differences but with augmented information from other related tables.
In industry, {\tt Google Trend} integrates an explanation component for single value and two relation;  \texttt{Tableau}~\cite{Tableau} provides {\tt Explain data}~\cite{Tableau19ExplainData} feature; \texttt{PowerBI}~\cite{Powerbi} supports functionality like {\tt Key Influener} ~\cite{PowerbiKeyInfluencers}; startups like \texttt{SisuData}~\cite{Sisu} is built to support ``why'' questions natively and scalably.
However, all prior works fall short of explaining aggregated time series because (1) only explaining one aggregated value overlooks the trend of time series -``why up/down”; (2) only focusing on two set differences (i.e., two endpoints in time series) dismisses the explanation in between.
Our \sys aims to identify the evolving explanations for aggregated time series over time.

\stitle{Time Series Segmentation}
Time series segmentation has been studied for decades. We note that the word “segmentation” is somewhat overloaded in the literature. One line of segmentation works focuses on visual-based piecewise linear approximation.
Specifically, ~\cite{koski1995syntactic, vullings1997ecg} use the sliding windows algorithm, which anchors the left point of a potential segment, then attempts to approximate the data to the right with increasing longer segments.
Douglas et al., ~\cite{douglas1973algorithms} and Ramer et al.,~\cite{ramer1972iterative} designs top-down algorithms to partition the visualization.
\cite{keogh1997probabilistic} and  \cite{hunter1999knowledge} have used the bottom-up algorithm to merge from the finest segments.
Keogh et al.  \cite{keogh2004segmenting} show that the bottom-up algorithm achieves the best results compared with sliding window and top-down, and further introduced a new online algorithm that combines the sliding window and bottom-up to avoid rescanning of the data when streaming.

Another line of work is semantic segmentation including FLUSS ~\cite{gharghabi2017matrix, gharghabi2019domain}, AutoPlait~\cite{matsubara2014autoplait}, NNSegment proposed in LimeSegment ~\cite{sivill2022limesegment}, which aims to divide a time series into
internally consistent subsequences, e.g., segmenting the heartbeat cycles when a person switches from running to walking.
Thus, these algorithms require an extra input called subsequence length, e.g., a heartbeat cycle. However, our task is to explain the trends in the aggregated time series (e.g., the trend of total covid cases) instead of explaining or identifying the periodic difference in one time-series instance. Hence, our segmentation does not rely on the period length.

To conclude, unlike all above, \sys is the first to segment time series based on segments' explanations.

\stitle{Time-series ML Model Explanation}
The time-series ML model takes a univariate or multivariate time series as input and outputs a prediction label. Unlike text or image models, there is limited literature on black box Time-series ML Model explainability.  FIT~\cite{tonekaboni2020went} is an explainability framework that defines the importance of each observation based on its contribution to the black box model’s distributional shift. Similarly, Rooke et al. ~\cite{rooke2021temporal} extend FIT into WinIT, which measures the effect on the distribution shift of groups of observations. Labaien et al.~\cite{labaien2020contrastive} finds the minimum perturbation required to change a black box model output. Recently, LimeSegment~\cite{sivill2022limesegment} selects representative input time series segments as explanations for time-series classifier's output.
In these works, the “explain target” is the prediction label and the time series serves as the “explain feature”. Contrarily, the “explain target” in \sys is the up/down trend in an aggregated time series and the explain-by attributes are our “explain features”. Unlike explaining instance-level prediction of the black box ML model, \sys is aggregation-level explanation for white-box aggregation.

%% file: content/probOverview.tex
\section{Problem Overview}\label{sec:prob}

In this section, we start with existing works on two-relations diff and how they fall short for explaining aggregated time series, followed by formal formulations of our problem: {\em K-Segmentation}.

\subsection{Background} \label{ssec:background}
\subsubsection{Two-Relations Diff.}~\label{sssec:two-relations}
Diff operator~\cite{Abuzaid2018DIFFAR} focuses on identifying the difference between two relations. Given a test relation $R_t$ and a control relation $R_c$, the diff operator returns {\em explanations} describing how these two relations differ.
\begin{definition} [Explanation~\cite{Abuzaid2018DIFFAR}] \label{def:explanation}
    \normalfont Given a set of explain-by attributes $\A$, an explanation $E$ of order $\beta$ is defined as a conjunction of $\beta$ predicates, denoted as $E=(A_{1}$=$a_{1} \&...\& A_{\beta}$=$a_{\beta})$ where $A_{i}\in\A$.
\end{definition}

Explain-by attributes $\A$ can be specified by users based on their domain knowledge; otherwise, dimension attributes from $R_t$ are used.
Intuitively, an explanation $E$ corresponds to a data slice satisfying the predicate, and this data slice contributes to the overall difference between $R_t$ and $R_c$.
To quantify how well an explanation $E$ explains the difference between $R_t$ and $R_c$, diff operator~\cite{Abuzaid2018DIFFAR} provides a {\em difference metric} abstraction, denoted as $\gamma(E)$. Such abstraction is capable of encapsulating the semantics of many prior explanation engines~\cite{Bailis2017MacroBasePA, Wu2013ScorpionEA, Wang2015DataXA}. Commonly used $\gamma(E)$ include \texttt{absolute-change}, \texttt{relative-change}, \texttt{risk-ratio}. Throughout this paper, we focus on \texttt{absolute-change}. Other metrics can be applied similarly.

\begin{definition} [\texttt{Absolute-change}] \label{def:absolute-change}
    \normalfont
    Given a test relation $R_t$, a control relation $R_c$, an aggregate function $f(M, R)$ on some measure attribute $M$ in relation $R$, and an explanation $E$, the absolute change refers to the absolute difference between  $[f(M, R_t) - f(M, R_c)]$ before and after removing records that explanation $E$ corresponds to, i.e., $\gamma(E) = |[f(M, R_t) - f(M, R_c)] - [f(M, R_t - \sigma_{\scaleto{E}{3.5pt}} R_t)) - f(M, R_c - \sigma_{\scaleto{E}{3.5pt}} R_c)]|$, where $\sigma_{\scaleto{E}{3.5pt}} R_t$ and $\sigma_{\scaleto{E}{3.5pt}} R_c$ denote records satisfying the predicate $E$ in relation $R_t$ and $R_c$ respectively.
\end{definition}

As the name indicates, \texttt{absolute-change} only cares about the absolute change of $[f(M, R_t) - f(M, R_c)]$ no matter the change is an increase or decrease. To distinguish an increase from a decrease, we use $\tau(E)$ to denote the change effect caused by including data that $E$ corresponds to: intuitively, if including $E$ leads to an increase in $[f(M, R_t) - f(M, R_c)]$, $\tau(E)=+$; otherwise, $\tau(E)=-$.

\begin{definition} [Change Effect] \label{def:effect}
    \normalfont
    Following the setting in Definition~\ref{def:absolute-change}, the change effect of an explanation $E$ is defined as $\tau(E)=sign([f(M, R_t) - f(M, R_c)] - [f(M, R_t - \sigma_{\scaleto{E}{3.5pt}} R_t)) - f(M, R_c - \sigma_{\scaleto{E}{3.5pt}} R_c)])$.
\end{definition}

Now we have described \texttt{Absolute-change} as an example of $\gamma(E)$. With a difference metric $\gamma(E)$, we can then rank each candidate explanation and return top-m explanations with the highest $\gamma(E)$. However, these top-m explanations may contain overlapping records. Consequently, the effects of these records get duplicated, introducing bias in the top-m explanations.
Alternatively, we can constrain these $m$ explanations to be non-overlapping and define {\em top-m non-overlapping explanations} as in Definition~\ref{def:top-m}.
Two explanations $E_1$ and $E_2$ are said to be {\em non-overlapping} if their correspondent records are non-overlapping in any relation $R$, i.e., $\sigma_{\scaleto{E_1}{4pt}} R \cap \sigma_{\scaleto{E_2}{4pt}} R = \emptyset, \forall R$.

\begin{table} 
    \begin{center}
        \begin{tabular}{c|c|c}
            \hline
            Symb.       & Definition                       & Mathematical Expression                                                               \\
            \hline
            \hline
            $\A$        & explain-by attributes            & $\A$=$\{A_1,A_2,...\}$                                                                \\
            \hline
            $E$         & \makecell{an explanation}        & $E$=$(A_{1}$=$a_{1}..\&A_{\beta}$=$a_{\beta}), A_i\in\A$                              \\
            \hline
            $\gamma(E)$ & difference score of $E$          & Definition~\ref{def:absolute-change}                                                  \\
            \hline
            $\tau(E)$   & change effect of $E$             & Definition~\ref{def:effect}                                                           \\
            \hline
            $\E_m$      & \makecell{m non-overlap $E$}     & $\E_m$=$\{E_1,...E_m\}$                                                               \\
            \hline
            $\E_m^*$    & \makecell{top-m non-overlap $E$} & $\E_m^*$=$\argmax_{\scaleto{\E_m}{4pt}} [\sum_{\scaleto{E\in\E_m}{3.5pt}} \gamma(E)]$ \\
            \hline
            $ts$        & \makecell{time series}           & $ts$=$\{p_1,..p_i,..p_n\}$ over $[t_1, t_n]$                                          \\
            \hline
            $c_i$       & $i^{th}$ cutting position        & $c_i\in[1, n]$ at time $t_{c_i}$                                                      \\
            \hline
            $P_i$       & \makecell{segment $i$}           & $P_i$=$[p_{c_i}, p_{c_{i+1}}]$ from $t_{c_i}$ to $t_{c_{i+1}}$                        \\
            \hline
            $\P_K$      & \makecell{$K$-segment scheme}    & $\P_K$=$\{P_1,P_2,..,P_K\}$                                                           \\
            \hline
            $\EE$       & \makecell{evolving explanations} & $\EE$=$[\E_m^*(t_{c_1}, t_{c_2}),..,\E_m^*(t_{c_{k}}, t_{c_{k+1}})]$                  \\
            \hline
        \end{tabular}
    \end{center}
    
    \caption{Notations}
\end{table}

\begin{definition} [m Non-Overlapping Explanations]\label{def:m-non-overlapping}
    \normalfont
    Given a difference metric $\gamma(E)$ and an explanation order threshold $\bar{\beta}$, let $\E_m$ denote a set of $m$ non-overlapping explanations, i.e., $\E_m=\{E_1,...E_m\}$, where each $E_i$ has its order $\leq \bar{\beta}$ and is non-overlapping with $E_j$, $\forall E_i, E_j \in \E_m$.
\end{definition}

\begin{definition} [Top-m Non-Overlapping Explanations~\footnote{Definition~\ref{def:top-m} is defined over $\E_m$. Alternatively, we can define top-m as {\em at most} $m$ explanations, i.e., $\E_m^*$=$\argmax_{\{\scaleto{\E_x}{4pt}|\scaleto{x\leq m}{4pt}\}} [\sum_{\scaleto{E\in\E_x}{3.5pt}} \gamma(E)]$. Our proposed solution in Section~\ref{sec:design} and \ref{sec:sys} can work with it in a similar way.}]\label{def:top-m}
    \normalfont
    Top-m non-overlapping explanations are defined as $\E_m$ with the highest accumulative diff score, i.e., $\E_m^*$=$\argmax_{\scaleto{\E_m}{4pt}} [\sum_{\scaleto{E\in\E_m}{3.5pt}} \gamma(E)]$.
\end{definition}

Cascading analyst algorithm~\cite{Ruhl2018TheCA} is designed for returning top-m non-overlapping explanations. We will use term {\em top-explanation} for simplicity whenever there is no ambiguity.

\begin{example}[Two-Relations Diff]
    Consider the two points $p_1$ and $p_2$ in Figure~\ref{fig:total-covid-us} --- the underlying data corresponds to $p_1$ and $p_2$ constitute a control relation $R_c$ and a test relation $R_t$ respectively. Two-relations diff aims to explain the difference between $R_c$ and $R_t$. First, users can specify a set of explain-by attributes, e.g., [{\small \texttt{state}, \texttt{County}}]. Take explanation $E$=({\small \texttt{state=CA}}) as an example. It is of order one, i.e., $\beta=1$ and its difference score $\gamma(E)$ can be calculated as $|(p_2.v-p_1.v)-(p_2'.v-p_1'.v)|$ using \texttt{absolute-change} as shown in Figure~\ref{fig:total-covid-us}.
    We can then obtain Top-3 non-overlapping explanations for differing $R_t$ and $R_c$ using cascading analyst algorithm~\cite{Ruhl2018TheCA} --- $\E_3^*$=$\{$$E_1$=({\small \texttt{state=CA}}), $E_2$=({\small \texttt{state=TX}}), $E_3$=({\small \texttt{state=FL}})$\}$.

\end{example}

\subsubsection{Aggregated Time Series.}
Time series is a series of data points indexed in time order. An {\em aggregated time series} refers to a special type of time series, where each point $p$ is associated with a timestamp $p.t$ and an aggregated value $p.v$, derived by aggregating all records at timestamp $p.t$. Essentially, an aggregated time series corresponds to the result of some group-by query. Consider a relation $R$ with $\{D_i\}$ dimension attributes and $\{M_j\}$ measure attributes, and a group-by query in the form of {\small {\{\tt SELECT $\T$, f(M) FROM R GROUP BY $\T$\}}}, where $\T$ denotes some time-related ordinal dimension ($\T\in \{D_i\}$), and $f(M)$ is some aggregate function on measure $M$ ($M\in \{M_j\}$). The query result can be denoted by an aggregated time series $\ts$ with value $f(M)$ over time dimension $T$.
\begin{definition} [Aggregated Time Series]
    \normalfont An aggregated time series $\ts$ over time $[t_1, t_n]$ is a series of points $\{p_1,...,p_i,...,p_n\}$ ordered by time dimension $T$ and each point's value $p_i.v$ is an aggregated number from a list of records with the same time $p_i.t$. %
\end{definition}

Data analysts often visualize aggregated time series to help understand data's overall trend as time goes along as shown in Figure~\ref{fig:total-covid-us}. A natural follow-up question is ``what makes ups and downs". Different from two-relations diff described in Section~\ref{sssec:two-relations}, this "explain" question focuses on the whole time horizon and the underlying top-explanation tends to evolve dynamically along the time even when the overall trend looks the same visually.

\begin{definition} [Evolving Explanations]
    \normalfont Given $m$ and an aggregated time series $\ts$ over time $[t_1, t_n]$, {\em evolving explanations} is a sequence of top-explanation at different periods, denoted as $\EE=[\E_m^*(t_{c_1}, t_{c_2}),\E_m^*(t_{c_2}, t_{c_3}),...,\E_m^*(t_{c_{k}}, t_{c_{k+1}})]$, where ${c_1}=1$, ${c_{k+1}}=n$, $\{c_2,c_3..c_{k}\}$ denote the $(k$-$1)$ cutting positions in between, and each $\E_m^*(t_{c_i}, t_{c_{i+1}})$ denotes top-explanation from $t_{c_i}$ to $t_{c_{i+1}}$.
\end{definition}

\begin{example} [Evolving Explanations] \label{exp:evolve-exp}
    Figure~\ref{fig:total-covid-us} depicts an aggregated time series that corresponds to a groupby-aggregate query with $f(M)$=\texttt{SUM(total\_confirmed\_cases)} on table \texttt{Covid-19}.
    Figure~\ref{fig:covid-explain} illustrates the underlying evolving explanations for the increase in Figure~\ref{fig:total-covid-us}. We have six different time periods, where each period shares the same intrinsic explanations while neighboring periods have different ones. For instance, the top-3 contributors are $\E_3^*(t_{c_2}, t_{c_3})$=$\{$$E_1$=({\small \texttt{state=NY}}), $E_2$=({\small \texttt{state=NJ}}), $E_3$=({\small \texttt{state=MA}})$\}$ during $t_{c_2}$=\texttt{2020-3-14} and $t_{c_3}$=\texttt{2020-5-4}; $\E_3^*(t_{c_6}, t_{c_7})$=$\{$$E_1$=({\small \texttt{state=CA}}), $E_2$=({\small \texttt{state=TX}}), $E_3$=({\small \texttt{state=FL}})$\}$ from \texttt{2020-11-27} to \texttt{2020-12-31}.
\end{example}

\subsection{Problem Definition} \label{ssec:prob_def}

Motivated by the observation that top contributor (i.e., explanation) evolves over time, we study the problem of identifying {\em evolving explanations} for the continuous changes happened in an aggregated time series.
The overall problem of identifying evolving explanations can be decomposed into two sub-problems: {\em (a).} segmentation; {\em (b).} find the explanations that contributes most in each segment.
These two sub-problems are intertwined  with each other: the goodness of a segmentation scheme depends on how cohesive top-explanations are within each segment; meanwhile, top-explanation are derived for each segment after obtaining the optimal segmentation scheme. We remark that performing segmentation without considering explanation information is insufficient, as we will demonstrate experimentally in Section~\ref{sec:exp}.

\stitle{Segmentation.}
To explain the continuous change in an aggregated time series $\ts$ over time $[t_1, t_n]$, we need to partition the whole time domain $[t_1, t_n]$ into non-overlapping segments, such that each segment $P_i$=$[p_{c_i}, p_{c_{i+1}}]$ during time $t_{c_i}$ and $t_{c_{i+1}}$ shares the same intrinsic explanations while neighboring segments have different ones. This resembles the well-studied clustering problem, whose goal is to minimize within-cluster variance and maximize inter-cluster variance. In particular, given a segment number $K$, we abstract our problem as a {\em K-Segmentation} problem, adapting the optimization formula from K-Means~\cite{K-means-optimization}. Let $\P_K$ denote a {\textit K-segmentation} scheme: $\P_K=\{P_1$=$[p_{c_1}, p_{c_2}], P_2$=$[p_{c_2}, p_{c_3}]...P_K$=$[p_{c_K}, p_{c_{K+1}}]\}$, where ${c_1}$=$1$, $c_{K+1}$=$n$, and $\{c_2,c_3...,c_K\}$ are $(K$-$1)$ cutting positions. In Example~\ref{exp:evolve-exp} (Figure~\ref{fig:covid-explain}), $K$=6 and the cutting positions $\{c_2,...,c_5\}$ correspond to time \{\texttt{3-14}, \texttt{5-4}, \texttt{5-29}, \texttt{9-25}, \texttt{11-27}\}. Next, we formulate {\em K-Segmentation} problem as below:

\begin{problem}[$K$-Segmentation] \label{prob:k-segmentation}
\normalfont Given an aggregated time series $\ts$ and a segmentation number $K$, identify the optimal segmentation scheme $\P_K^*=\argmin_{\P_K=\{P_1,...,P_K\}} \sum_{i=1}^K$$|P_i|\var(P_i)$ where $\var(P_i)$ denotes the variance in segment $P_i$.
\end{problem}

We remark that the design of within-segment variance $\var(P_i)$ is critical to the effectiveness of $K$-Segmentation. No prior works have studied $\var(P_i)$ with the goal of quantifying explanation consistency. This is challenging as we will dive into in Section~\ref{sec:design}.

\stitle{Explain trend in each segment.}
Given a fixed segment $P_i=[p_{c_i}, p_{c_{i+1}}]$ from time $t_{c_i}$ to $t_{c_{i+1}}$, we will now describe how to explain the trend in $P_i$ with only one static top-explanation $\E_m^*(t_{c_i}, t_{c_{i+1}})$ --- static top-explanation is a special case of evolving explanations $\EE$ with segment number $K=1$.
If $P_i$ is cohesive, meaning that $P_i$ has consistent top-explanation during $t_{c_i}$ and $t_{c_{i+1}}$, we can simply focus on its two endpoints and then employ prior works on two-relations diff (Section~\ref{ssec:background}). The derived top-explanation $\E_m^*$ explains the changes from time $t_{c_i}$ to $t_{c_{i+1}}$. However, when $P_i$ is not cohesive, there exists no single static top-explanation $\E_m^*$ that is capable of explaining the whole trend evolvement in $P_i$. Nevertheless, we can still derive some static top-explanation by looking at its two endpoints and using two-relations diff~\cite{Ruhl2018TheCA}, though the explanation quality might be poor. As we will elaborate in Section~\ref{ssec:design-var}, each segment in the optimal K-segmentation $\P_K^*$ is deemed to be cohesive, and the case of incohesive segment would only occur during the exploration phase over candidate segmentation schemes.
In all, when given a segment $P_i=[p_{c_i}, p_{c_{i+1}}]$, we first obtain a control relation $R_c$={\small {\{\tt SELECT * FROM R WHERE $\T=t_{c_i}$\}}} and a test relation $R_t$={\small {\{\tt SELECT * FROM R WHERE $\T=t_{c_{i+1}}$\}}} at two endpoints, and then derive top-explanation with cascading analyst algorithm~\cite{Ruhl2018TheCA}.

Now that we can exploit existing works for deriving static top-explanation within each segment, our problem of identifying evolving explanations boils down to a K-segmentation problem. As $K$ is not easy to specify in practice, \sys identifies the optimal $K$ for users by default, as we will discuss in Section~\ref{sec:optimal-k}.

%% file: content/design.tex
\section{K-Segmentation} \label{sec:design}
Designing a good within-segment variance to quantify explanation consistency is the key to our work's success and it is definitely non-trivial. In this section, we will start with our design of $\var(P_i)$ in Problem~\ref{prob:k-segmentation}, followed by an experiment illustrating its effectiveness.

\subsection{Design of Within-Segment Variance}\label{ssec:design-var}

Per our definition in \Cref{prob:k-segmentation}, K-Segmentation is similar to K-Means clustering. We carefully design the within-segment variance $\var(P_i)$ through making analogy to K-Means algorithm. However, different from K-Means or any existing variance design, $\var(P_i)$ in K-Segmentation should be regarding the variance of top-explanations within each segment $P_i$.

First, let us review the problem formulation of K-Means. Given a set of {\em objects} $(o_1,o_2,...,o_n)$ as inputs, where each object is a $d$-dimensional vector, K-Means aims to partition these $n$ objects into $K$ partitions $\P_K$=$\{P_1,P_2..P_K\}$ minimizing the within-cluster variance:

\begin{equation}\label{eq:k-means-obj}
    \argmin_{\P_K} \sum_{i=1}^K |P_i|\var(P_i)
\end{equation}

\begin{equation}\label{eq:k-means-variance}
    \var(P_i) = \frac{1}{|P_i|}\sum_{o\in P_i} dist(o, \mu_i)
\end{equation}

where $\mu_i$ is the centroid of partition $P_i$ and $dist(o, \mu_i)$ denotes the distance between an object $o$ and the centroid $\mu_i$ in $P_i$, e.g., $L2$ distance.
Comparing Problem~\ref{prob:k-segmentation} with Eq.~\ref{eq:k-means-obj}, we can see K-Segmentation employs the same optimization objective as K-Means but with a customized $\var(P_i)$. To develop a good $\var(P_i)$ in K-Segmentation, careful thoughts are required around: {\em (1)} what is an ``object''; {\em (2)} what is the centroid of a segment; and {\em (3)} how to measure the distance between object and centroid based on explanations.

\subsubsection{Object in K-Segmentation.}
\begin{figure}
    \centering
    \includegraphics[width=0.8\columnwidth]{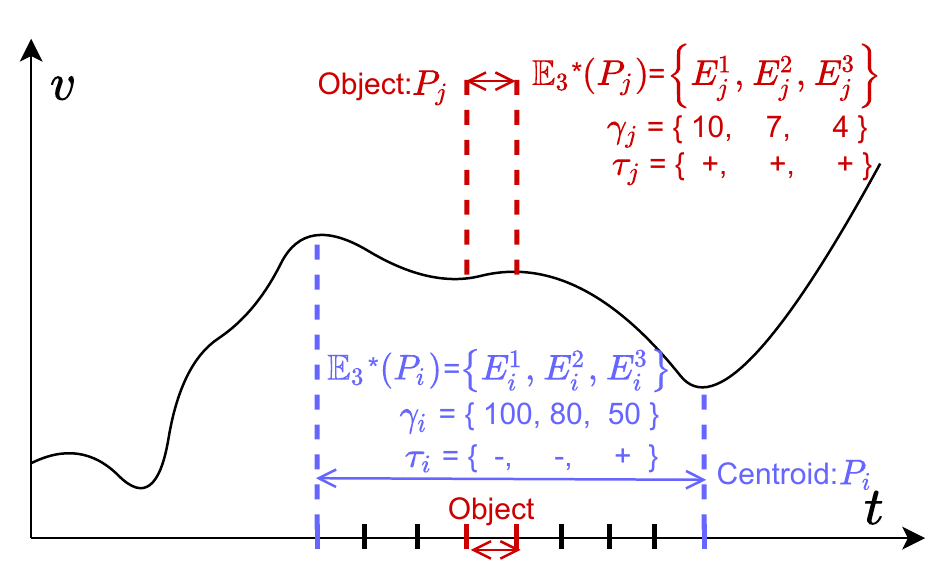}
    \caption{Object, Centroid, and Top-Explanations}
    \label{fig:distance}
\end{figure}

Given an aggregated time series $\ts=\{p_1, p_2, ..., p_n\}$, K-Segmentation aims to segment $ts$ into $K$ partitions such that explanations are shared within each partition. Since each single point itself cannot reveal any time series trend, the atomic unit for partitioning is a segment of size two, i.e., $[p_i, p_{i+1}]$. That is, an object in K-Segmentation refers to a segment of size two as shown in Figure~\ref{fig:distance} and there is in total $(n-1)$ objects, i.e., $\{o_1=[p_1, p_2], o_2=[p_2,p_3],... , o_{n-1}=[p_{n-1}, p_n]\}$.

\subsubsection{Centroid of a partition.}
Different from the general K-Means, objects in K-Segmentation follows a time ordering where $o_1<o_2<...<o_{n-1}$, and a partitioning scheme in K-Segmentation is only valid when objects in each partition form a segment. Hence, a partition in K-Segmentation can be denoted as $P_i=[p_{c_i}, p_{c_{i+1}}]$ with objects $\{o_{c_i}$=$[p_{c_i}, p_{c_i+1}], o_{c_i+1}..,o_{c_{i+1} -1}$=$[p_{c_{i+1}-1}, p_{c_{i+1}}]\}$. Naturally, we can use segment $[p_{c_i}, p_{c_{i+1}}]$ as the centroid of partition $P_i$.

\subsubsection{Distance between object and centroid.}
In essence, both object and centroid are segments of the input time series $\ts$. Next, we focus on the design of distance between segments. Distance between two objects or between an object and a centroid follow naturally.

\stitle{High-Level Idea.}
As our goal is to group objects with the same top-explanations into one partition, the distance between two segments shall be based on their top-explanations. Given two segments $P_i$ and $P_j$, their top-explanations $\E_m^*(P_i)$ and $\E_m^*(P_j)$ can be derived based on Section~\ref{ssec:background}. Each is a ranked list of explanations, i.e., $\E_m^*(P_i)=[E_{i}^1, E_{i}^2,...,E_{i}^m]$ and $\E_m^*(P_j)=[E_j^1, E_j^2,...,E_j^m]$ as shown in Figure~\ref{fig:distance}. Strawman approaches like measuring the Jaccard distance between $\E_m^*(P_i)$ and $\E_m^*(P_j)$ fall short when there are multiple explain-by attributes. For instance, it is unclear how to quantify the partial overlap between $E_i^1$=$\texttt{\{state=WA\}}$ and $E_j^2$=$\texttt{\{state=WA and age>50\}}$. Alternatively, we can measure the distance between $P_i$ and $P_j$ by how well $\E_m^*(P_i)$ explains $P_j$ and how well $\E_m^*(P_j)$ explains $P_i$.

\stitle{How well $\E_m^*(P_j)$ explains $P_i$.}
We draw inspirations from information retrieval community. Normalized discounted cumulative gain (NDCG) is a commonly used measure for ranking quality in information retrieval and we adapt NDCG to quantify how well $\E_m^*(P_j)$ explains $P_i$.
To model our scenario after the web search setting, we can treat each segment $P_i$ as a query and each explanation $E$ as a document. $\E_m^*(P_j)$ corresponds a ranked list of retrieved documents returned by the search engine, while $\E_m^*(P_i)$ is the ideal retrieved document list. The relevance between a segment $P_i$ and an explanation $E$ is quantified by the difference metric $\gamma(E, P_i)$. As illustrated in Table~\ref{table:DCG} (row in blue), given an explanation $E_j^r\in \E_m^*(P_j)$ with rank $r$, the relevance of $E_j^r$ towards $P_i$ can be calculated as $\gamma(E_j^r, P_i)$.
However, explanation $E_j^r$ might make KPI increase in segment $P_i$, but lead to a decrease in segment $P_j$ (see Definition~\ref{def:effect}). Thus, when $E_j^r$ has opposite effects on $P_i$ and $P_j$, we shall rectify the relevance to zero as our ultimate goal is to measure the distance between $P_i$ with $P_j$. Formally, we denote the {\em rectified relevance} as $\bar{\gamma}(E_j^r, P_i)$ and we have $\bar{\gamma}(E_j^r, P_i)$=$\gamma(E_j^r, P_i) \times \mathbbm{1}_{\tau(E_j^r, P_j)=\tau(E_j^r, P_i)}$.
Take $E_j^3$ in Table~\ref{table:DCG} as an example, $E_j^3$ contributes the increase in segment $P_j$ but the decrease in $P_i$, thus the relevance is rectified to 0.

\begin{table}
    \begin{center}
        \small
        \begin{tabular}{ |c|c|c|c||c|c| }
            \hline
                                  &                       & \multicolumn{4}{c|}{How well $\E_m^*(P_j)$ explains $P_i$}                                                                                                                              \\
            \hline
            \multirow{2}{*}{Rank} & \multirow{2}{*}{Expl} & \multicolumn{2}{c|}{Effect (+/-)}                          & Relevance                        & Rectified                                                                               \\
            \cline{3-4}
                                  &                       & on $P_j$                                                   & on $P_i$                         & on $P_i$                           & Relevance $\bar{\gamma}$                           \\
            \hline
            \hline
            \color{blue}{r}       & \color{blue}{$E_j^r$} & \color{blue}{$\tau(E_j^r, P_j)$}                           & \color{blue}{$\tau(E_j^r, P_i)$} & \color{blue}{$\gamma(E_j^r, P_i)$} & \color{blue}{\makecell{$\gamma(E_j^r, P_i) \times$ \\ $\mathbbm{1}_{\tau(E_j^r, P_j)=\tau(E_j^r, P_i)}$}}  \\
            \hline
            \hline
            1                     & $E_j^1$               & +                                                          & +                                & $\gamma(E_j^1, P_i)$               & $\gamma(E_j^1, P_i)$                               \\
            \hline
            2                     & $E_j^2$               & +                                                          & +                                & $\gamma(E_j^2, P_i)$               & $\gamma(E_j^2, P_i)$                               \\
            \hline
            3                     & $E_j^3$               & +                                                          & -                                & $\gamma(E_j^3, P_i)$               & 0                                                  \\
            \hline
            \hline
            \multicolumn{6}{|c|}{\multirow{2}{*}{Discounted cumulative gain (DCG):\hspace{4mm} $\frac{\gamma(E_j^1, P_i)}{\log_2(1+1)}+\frac{\gamma(E_j^2, P_i)}{\log_2(1+2)} + \frac{0}{\log_2(1+3)}$}}                                            \\
            \multicolumn{6}{|c|}{}                                                                                                                                             \\
            \hline
        \end{tabular}
    \end{center}
    \caption{Example of DCG Between $\E_m^*(P_j)$ and $P_i$ }
    \label{table:DCG}
\end{table}

Now we have mapped our scenario to query-document retrieval setting, i.e., query $P_i$, retrieved document list $\E_m^*(P_j)$, and the rectified relevance formula $\bar{\gamma}$, $NDCG(P_i, \E_m^*(P_j))$ can be calculated using Eq.~\ref{eqn:dcg}, \ref{eqn:idcg}, and \ref{eqn:ndcg}. It quantifies how well $\E_m^*(P_j)$ explains $P_i$, with ranges from 0 to 1. At an extreme, when $\E_m^*(P_j)$ is exactly the same as $\E_m^*(P_i)$ and with the same effect on $P_i$ and $P_j$, $NDCG(P_i, \E_m^*(P_j))$=1, meaning $\E_m^*(P_j)$ explains $P_i$ perfectly.

\begin{equation} \label{eqn:dcg}
    DCG(P_i, \E_m^*(P_j)) = \sum_{r=1}^m \frac{\bar{\gamma}(E_j^r, P_i)}{\log_2(r+1)}
\end{equation}
\begin{equation} \label{eqn:idcg}
    DCG(P_i, \E_m^*(P_i)) = \sum_{r=1}^m \frac{\bar{\gamma}(E_i^r, P_i)}{\log_2(r+1)} = \sum_{r=1}^m \frac{\gamma(E_i^r, P_i)}{\log_2(r+1)}
\end{equation}
\begin{equation} \label{eqn:ndcg}
    NDCG(P_i, \E_m^*(P_j)) = \frac{DCG(P_i, \E_m^*(P_j))}{DCG(P_i, \E_m^*(P_i))}
\end{equation}

\stitle{Calculating Distance.}
We can then define the distance between $P_i$ and $P_j$ as in Eq.~\ref{eqn:distance}, where $NDCG(P_i, \E_m^*(P_j))$ quantifies how well $\E_m^*(P_j)$ explains $P_i$ and $NDCG(P_j, \E_m^*(P_i))$ quantifies how well $\E_m^*(P_i)$ explains $P_j$.

\begin{equation} \label{eqn:distance}
    dist(P_i, P_j) = 1- \frac{NDCG(P_i, \E_m^*(P_j)) + NDCG(P_j, \E_m^*(P_i))}{2}
\end{equation}

Eq.~\ref{eqn:distance} averages $NDCG(P_i, \E_m^*(P_j))$ and $NDCG(P_j, \E_m^*(P_i))$ to obtain the similarity between $P_i$ and $P_j$, followed by a complement to get the distance. $dist(P_i, P_j)$ is symmetric with ranges $[0, 1]$.

\subsubsection{Putting all together.} Given a partition $P_i=[p_{c_i}, p_{c_{i+1}}]$, it contains a continuous list of objects $\{P_x=[p_{x}, p_{x+1}]\}$ where $c_i\leq x < c_{i+1}$ and its centroid is $P_i=[p_{c_i}, p_{c_{i+1}}]$. Using Eq.~\ref{eqn:distance} and \ref{eq:k-means-variance}, we can derive our variance of $P_i$ as in Eq.~\ref{eqn:variance}.

\begin{equation} \label{eqn:variance}
    \var(P_i) = \frac{1}{c_{i+1}-c_i}\sum_{x=c_i}^{c_{i+1}-1} dist(P_x, P_i), \texttt{ where } P_x=[p_{x}, p_{x+1}]
\end{equation}

Our problem formulation of K-Segmentation is now complete, by instantiating $\var(P_i)$ in Problem~\ref{prob:k-segmentation} with Eq.~\ref{eqn:variance}.

\input{content/effectiveness-variance.tex}

%% file: content/effectiveness-variance.tex
\subsection{Effectiveness of Variance} \label{ssec:exp-variance}
In this subsection, we evaluate the effectiveness of our variance design. We term our variance metric in \Cref{eqn:variance} {\tt tse}. Since real-world datasets lack its ground truth of evolving explanations, we synthesize datasets with ground truths and evaluate how {\tt tse} performs compared with other alternatives. We will evaluate the end-to-end effectiveness in \Cref{sec:exp}.

\subsubsection{Synthetic datasets}
\label{sec:syn}
We synthesize datasets and generate their ground truth $K$-Segmentation $\P_K^*$.
Each dataset is one relation $R$ with schema: \texttt{$\T$}, \texttt{sales}, \texttt{category}.
The aggregated time series represents how the total sales changes along the time \texttt{$\T$} --- {\small {\tt SELECT $\T$, count(sales) FROM R GROUP BY $\T$}}.
We set the explain-by attributes $\A$=$\{{\tt category}\}$ and there are three categories: \textit{$a_1$}, \textit{$a_2$}, \textit{$a_3$}. Each predicate e.g., {\tt \{category=$a_1$\}} denotes an explanation $E$.

\begin{figure}[h]
        \centering
        \begin{subfigure}{0.47\columnwidth}
            \centering
            \includegraphics[width=\columnwidth]{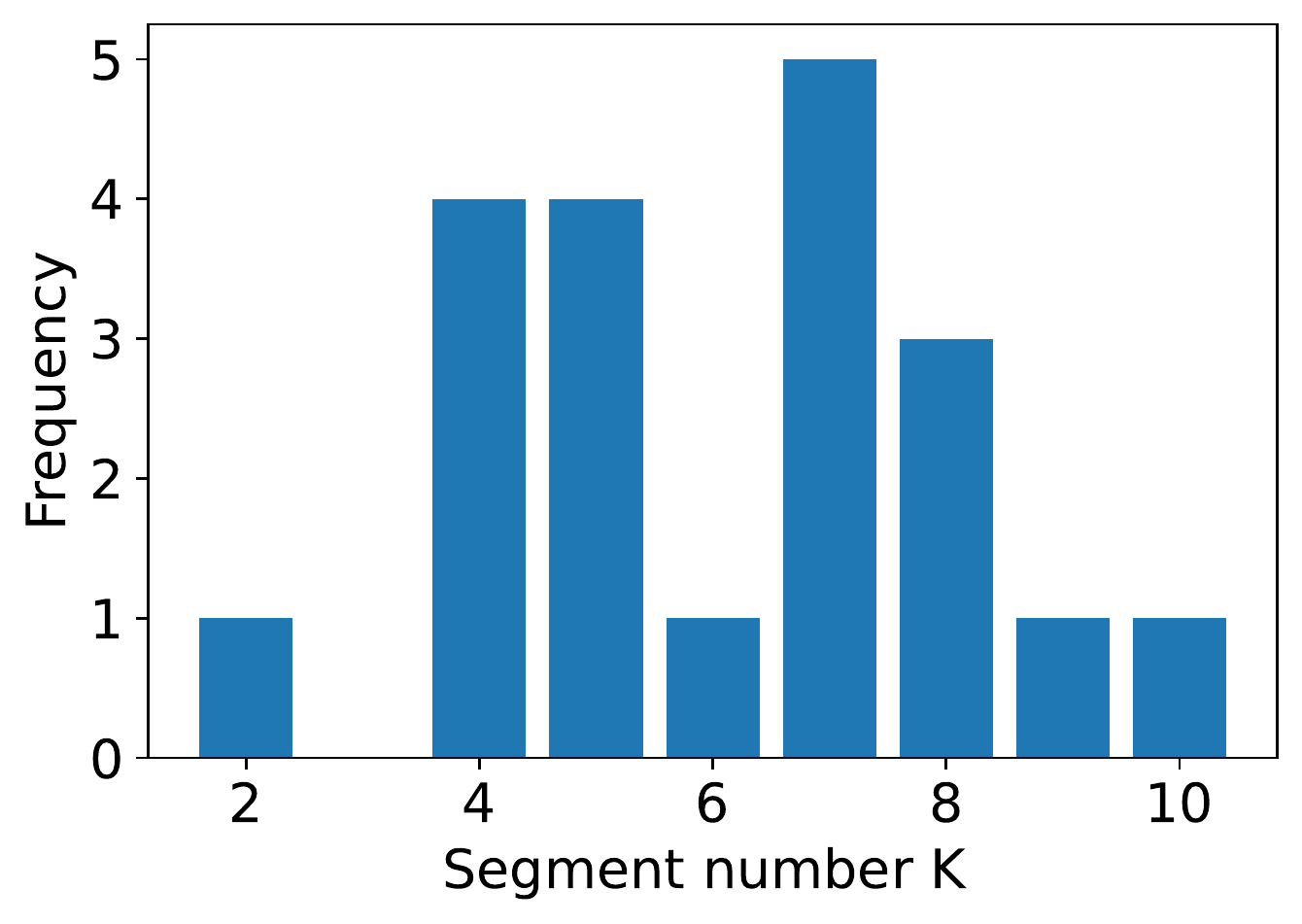}
        \end{subfigure}
        \begin{subfigure}{0.48\columnwidth}
            \centering
            \includegraphics[width=\columnwidth]{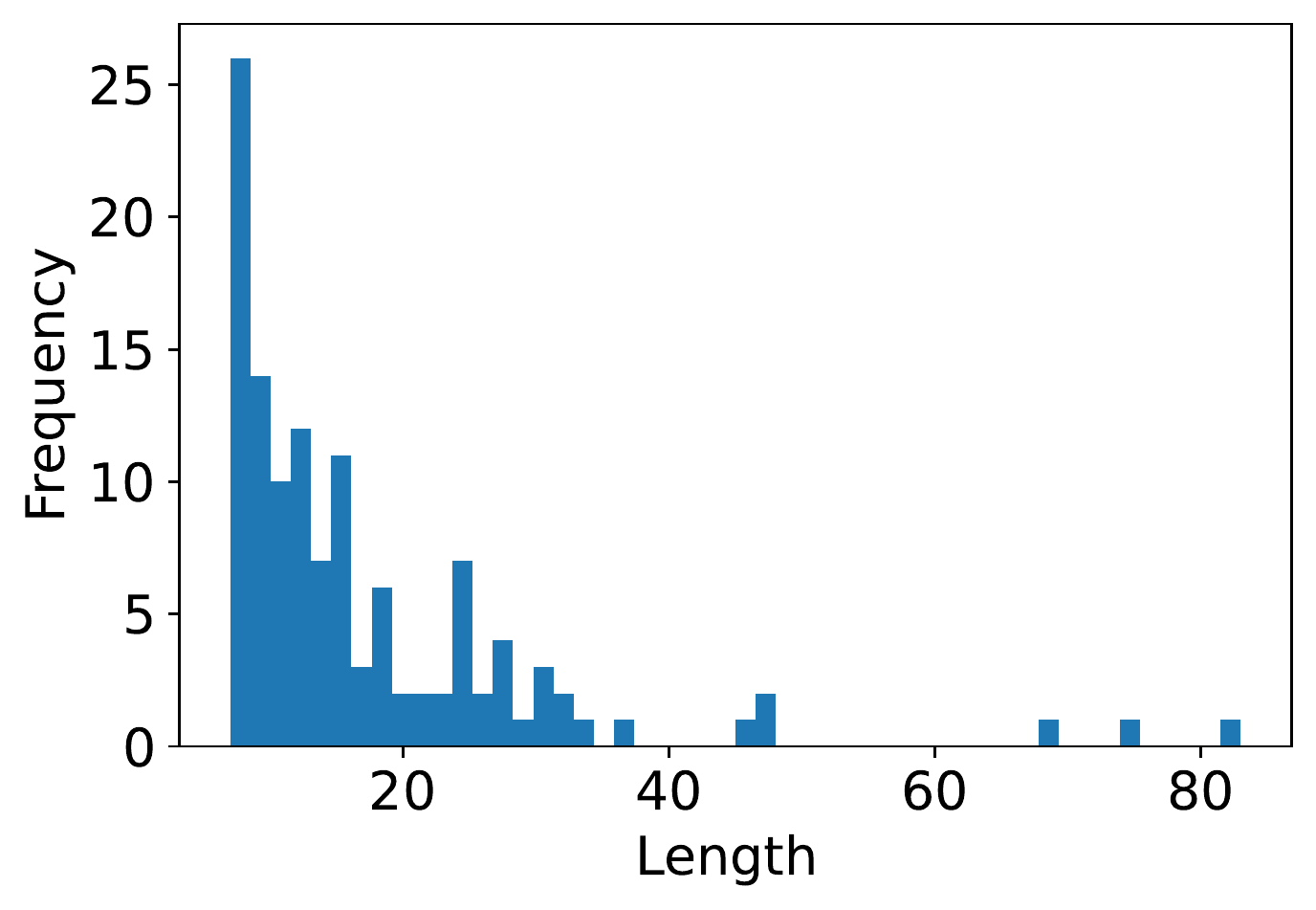}
        \end{subfigure}
        \caption{Distribution of segment number K and length of each segment.}
        \label{fig:dist}
    \end{figure}

\stitle{Synthesize Procedure}
The aggregated {\small {\tt sales}} time series can be viewed as a summation of each {\small {\tt category}}'s time series.
We start by synthesizing each category's time series. In detail, we first randomly pick cutting points $\{c_1^i,...,c_{j_i}^i\}$ for each category $E_i$'s time series --- {\small {\{\tt SELECT $\T$, count(sales) FROM R WHERE {\tt category=$a_i$} GROUP BY  $\T$\}}}. For each segment $[p_{c_{k}^i}, p_{c_{k+1}^i}], 1 \le k < j_i$ defined by these cutting points, we synthesize either an upward or downward trend in linear shape. We restrict the adjacent segments to have different up or down trends.
We can then derive the ground truth segmentation's cutting points of the aggregated time series as the union of each category's cutting  points, i.e., $\bigcup_{i=1}^{3} \{c_{1}^i,...,c_{j_i}^i \}$. We treat $\bigcup_{i=1}^{3} \{c_{1}^i,...,c_{j_i}^i \}$ as our ground truth because (1) each predicate has a consistent up or down trend in each segment; (2) our restriction that adjacent segments have different trend direction guarantees that every cutting point is necessary and $\bigcup_{i=1}^{3} \{c_{1}^i,...,c_{j_i}^i \}$ is the minimal coherent segmentation method. 
We set the time series' length at 100 and synthesize 20 datasets with seven different levels of $SNR_{dB}$. The lower the $SNR_{dB}$ is, the noisier the time series is. 
    We remark that the number K and the length of each segment are diverse in our synthetic datasets, with segment number K varying from 2 to 10 and segment length varying from 6 to 84 as shown in \Cref{fig:dist}.

\stitle{Signal-to-Noise Ratio}
Real-world data is quite noisy.  For our synthetic dataset, we add Gaussian Noise $N(0, \sigma^2)$ to each predicate's time series to simulate noisy time series. We quantify the noise level using signal-to-noise level, namely SNR~\cite{SNR}.
We add different noise  $SNR_{dB} = 20, 25..,50$ to each dataset.
The lower the $SNR$ is, the noisier the time series is.

\begin{figure}[h]
    \includegraphics[width=\columnwidth]{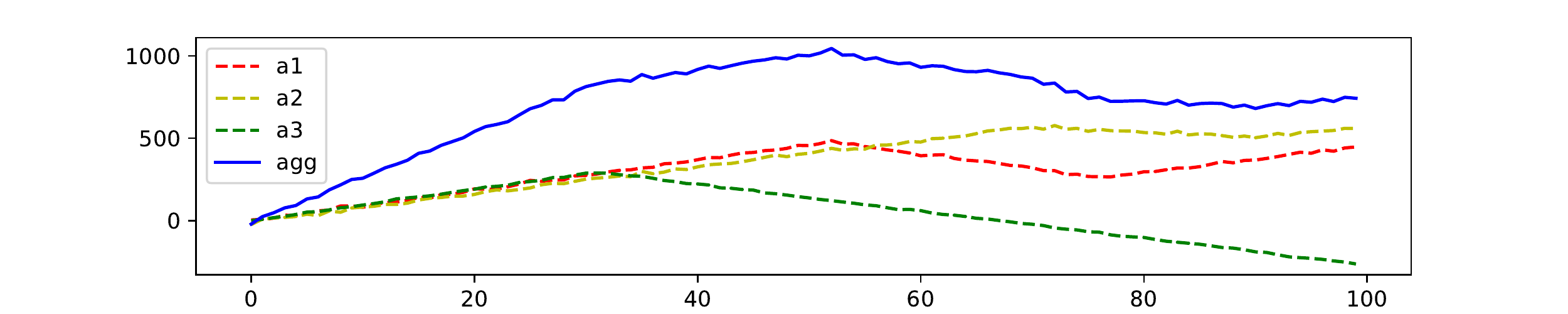}
    \caption{A synthetic time series with SNR = 35.}
    \label{linear-syn}
\end{figure}

\begin{example}[Synthetic Dataset and Ground Truth Segmentation]
In \Cref{linear-syn}, the dash lines illustrate each predicate's time series, and the noise level is SNR = 35. The predicate {\tt category = a1} has its cutting points at 52, 76, the predicate {\tt category = a2} has its cutting point at 70, 90, and the predicate{\tt category = a3} has its cutting point at 31. Thus, in this example, we can derive the aggregated time series cutting point as the union of three predicates' –  \{ 31, 52, 70, 76, 90 \}.
\end{example}

\subsubsection{Effectiveness}
We compare {\tt tse} with other alternatives in terms of effectiveness.

\stitle{Alternatives}
We design our alternatives by enumerating other reasonable distance metrics and variance structures.

    {\it Change $dist(P_i, P_j)$ definition and keep the variance structure.} Different from our metric in \cref{eqn:distance}, if we only consider how well each object's explanation $\E_m^*(P_j)$ explains the centroid $P_i$, we can derive \texttt{dist1} in \cref{eqn:sim1}; or if we only consider how well the centroid's  $\E_m^*(P_i)$ explains each object $P_j$, we can derive \texttt{dist2} in \cref{eqn:sim2}.

\begin{equation} \label{eqn:sim1}
    \texttt{dist1:} dist(P_i, P_j) = 1- NDCG(P_i, \E_m^*(P_j))
\end{equation}
\begin{equation} \label{eqn:sim2}
    \texttt{dist2:}  dist(P_i, P_j) = 1- NDCG(P_j, \E_m^*(P_i))
\end{equation}

{\it Change the variance structure and keep the $dist(P_i, P_j)$ definition.}  Instead of comparing the centroid with each object, we compare all possible object pairs. We name this metric \texttt{allpair}.

\begin{align} \label{eqn:all_pair_variance}
    \texttt{allpair:} \var(P_j) =AVG \{\sum_{P_x} \sum_{P_y} & dist(P_x, P_y)\}, \texttt{} ~  P_x, P_y \subseteq P_j
\end{align}

Based on {\tt tse, dist1, dist2, allpair}, if we further change the second term in the distance metric to its l2 norm, we can derive \texttt{Stse/Sallpair}, \texttt{Sdist1} and \texttt{Sdist2} correspondingly.
Till now, we have eight different forms of metrics.

\stitle{Evaluation}
As shown in \cref{eq:k-means-obj}, our problem is to find the segmentation that minimizes the objective score. Thus, the effectiveness criteria of variance design is whether the ground truth can score the lowest or close to the lowest in noisy datasets.

\begin{figure}[h] 
    \includegraphics[width=\columnwidth]{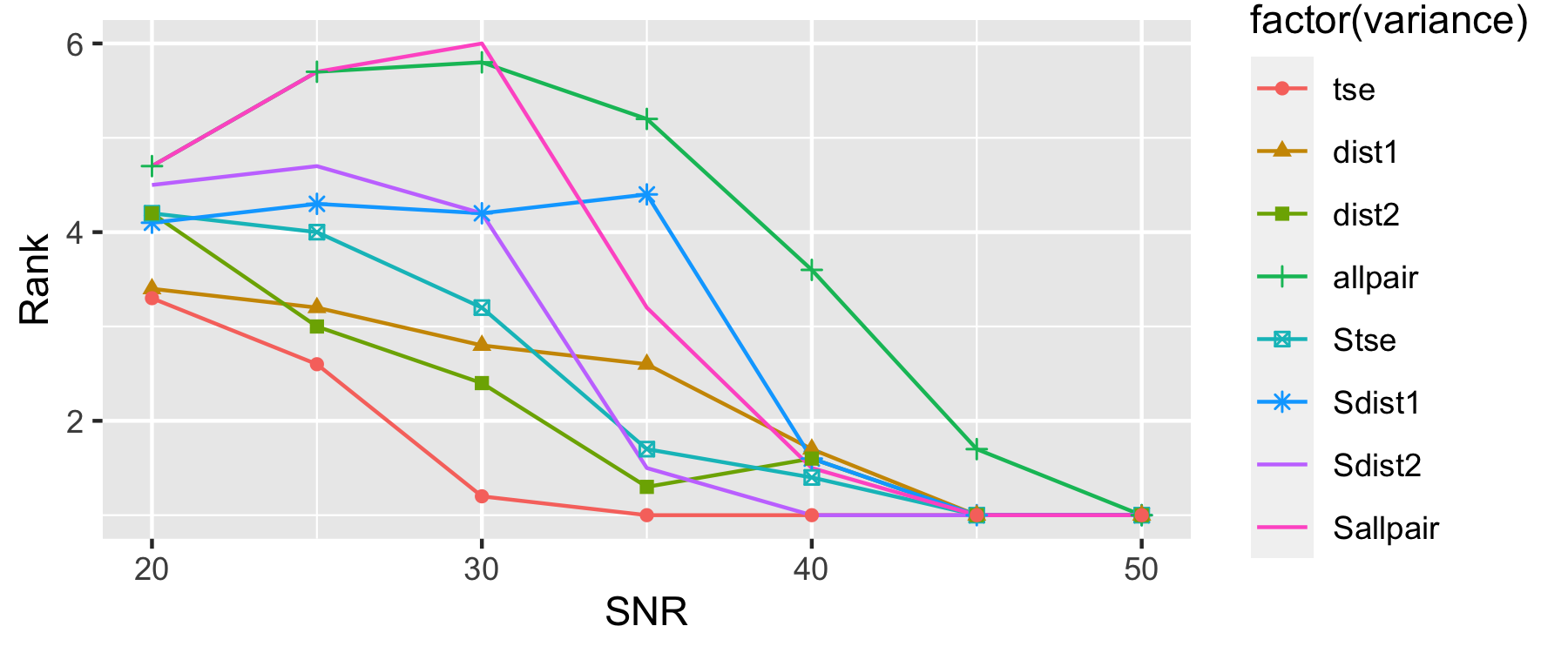}
    \caption{The {\it average rank} of all metrics at different SNRs.}
    \label{fig:rank}
\end{figure}

We compare {\tt tse} with all other alternatives. Given one metric and one dataset $D$ with its ground truth, the K-segmentation search space $\P_K$ is huge and each segmentation scheme has its metric score.
We study how the ground truth segmentation's score ranks among all possible segmentation schemes: the higher the rank is, the better the metric is. Because $\P_K$ space is huge, we sample 10000 random segmentation schemes, and rank the ground truth among all the samples, denoted as {\it ground truth rank}. On this dataset $D$, we calculate all eight metric's {\it ground truth rank} following the same method. To compare all the metrics on $D$, we rank across all the eight metrics from {\it rank} 1 (highest rank) to {\it rank} 8 (lowest rank) ascendingly based on their own {\it ground truth rank}. The higher a metric's  {\it rank} is, the fewer segmentation schemes have lower variance than the ground truth, indicating that this metric is more effective on $D$.
\Cref{fig:rank} averages each metric's {\it rank} over all datasets at the same SNR level and shows how different metrics' {\it ranks} change along with the noisy level.
When SNR = 50dB, we can see that all metrics rank 1st because the ground truth all achieves the lowest score.  What's more, \texttt{tse} metric always has the highest {\it rank} compared with other metrics no matter what level the SNR is.

\stitle{Takeaway.} Compared with other alternative metrics, {\tt tse} is the most effective one.

%% file: content/efficiency.tex
\section{Our Solution: \sys} \label{sec:sys}
\begin{figure}
                \centering
                \includegraphics[width=\columnwidth]{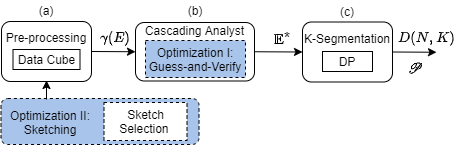}
                \caption{Solution Pipeline in \sys}
                \label{fig:pipeline}
            \end{figure}

Now we have formulated our K-Segmentation problem (Problem~\ref{prob:k-segmentation}), together with the designed within-segment variance in Eq.~\ref{eqn:variance}. In this section, we will present our solution \sys: we will start with a dynamic programming (DP) algorithm for solving K-Segmentation, assuming $\var(P_k)$ is available for each partition $P_k$; next, we will describe our solution pipeline (Figure~\ref{fig:pipeline}) -- steps for preparing $\var(P_k)$ and DP -- together with complexity analysis; last, we propose several optimizations for speeding up \sys.

\subsection{DP for K-Segmentation}\label{ssec:k-segmentation}

Different from K-Means, which is computationally intractable (NP-Hard), K-Segmentation is polynomial solvable. At a high-level, the search space in K-Means is $K^n$, while it is $\binom {n-2}{K-1}$ in K-Segmentation as the task is essentially to identify $(K-1)$ cutting positions among the $(n-2)$ non-endpoints of a given time series $\ts$.%

Intuitively, K-segmentation exhibits optimal substructure --- the optimal solution of K-segmentation can be constructed from the optimal solutions of its subproblems. Thus, we can use dynamic programming for solving K-Segmentation.
Let $D(j,k)$ denote the minimal total within-segment variance of $k$-segments over time range $[t_1, t_j]$, i.e., $D(j,k)=\min_{\P_k=[P_1,...,P_k]} \sum_{i=1}^k$$|P_i|\var(P_i)$. To derive $D(j,k)$, we can enumerate different positions of the last cut $j'$ and calculate the smallest total variance among all possible $j'$. When the last cut is fixed at position $j'$, the minimal total variance of $k$-segments can be decomposed into two parts: the minimal total variance of $(k-1)$-segments over time range $[t_1, t_{j'}]$ and the variance of the $k^{th}$ segment during $[t_{j'},t_j]$, i.e., $D(j', k-1) + |P_k|\var(P_k)$. The DP recursion function is expressed in Eq.~\ref{eq:k_segmentation_dp}.
            \begin{equation} \label{eq:k_segmentation_dp}
                \small
                D(j, k) = \argmin_{1 \le j' \le j }  [ D(j', k-1) + |P_k|\var(P_k) ] \texttt{,~~}~~~~ P_k=[p_{j'}, p_j]
            \end{equation}

            \underline{$D(j, k)$} is recursively computed with Eq.~\ref{eq:k_segmentation_dp}, which involves calculating \underline{variance $\var(P_k)$} for all segment $P_k=[p_{j'}, p_j]$ where $1\le j'< j \le n$. Given a segment $P_k=[p_{j'}, p_j]$, its $\var(P_k)$ is computed via the \underline{distance $dist(P_k, P_x)$} between centroid segment $P_k$ and each object segment $P_x=[p_x, p_{x+1}]$ where $j'\le x < j$ as shown in Eq.~\ref{eqn:variance}. Now to calculate $dist(P_k, P_x)$, we need to identify the \underline{top-explanations $\E_m^*$} in segment $P_j$ and $P_x$. Top-explanations $\E_m^*$ is derived with Cascading Analyst algorithm~\cite{Ruhl2018TheCA}, which requires computing the \underline{diff score $\gamma(E)$} for each explanation $E$.
            In Section~\ref{ssec:pipeline}, we will present our solution pipeline consisting of three modules: {\em (a.)} {\em Preprocessing module} for calculating $\gamma(E)$; {\em (b.)} {\em Cascading Analyst module} for deriving $\E_m^*$; and {\em (c.)} {\em K-Segmentation} module for computing $dist(P_k, P_x)$, $\var(P_k)$, and finally $D(j,k)$. We will also analyze the time complexity for each module, pinpointing the bottleneck in the whole pipeline for optimizations in Section~\ref{ssec:opt}.

            \subsection{Solution Pipeline and Complexity Analysis} \label{ssec:pipeline}
            The overall solution pipeline is depicted in Figure~\ref{fig:pipeline} with module (a) Preprocessing, (b) Cascading Analyst, and (c) K-Segmentation. In the following, we will dive into each module and analyze its computational complexity. The optimization modules with blue background in Figure~\ref{fig:pipeline} will be introduced in Section~\ref{ssec:opt}.

            \stitle{Precomputation.}
            Module (a) is responsible for computing the diff score $\gamma(E)$ for each candidate explanation $E$ over every segment $[p_{j'}, p_j]$ where $1\le j' < j \le n$. Given an explanation order threshold $\bar{\beta}$, we can enumerate all candidate explanations --- each is in the form of $E=[A_{1}$=$a_{1},...,A_{\beta}$=$a_{\beta}]$ where $1\le \beta\le \bar{\beta}$. Let $\epsilon$ be the total number of candidate explanations. By default, $\bar{\beta}$ is set as 3.

            As illustrated in Figure~\ref{fig:pipeline}, given an explanation $E$ and a segment $[p_{j'}, p_j]$, the difference score $\gamma(E)$, i.e., \texttt{absolute-change} in Definition~\ref{def:absolute-change}, can be calculated by looking at the points at time $t_{j'}$ and $t_j$ of the aggregated time-series $\ts(R)$ and $ts(R-\sigma_{\scaleto{E}{3.5pt}}R)$, i.e., the aggregated time-series from relation $R$ when excluding data with predicate $E$. As most aggregate function $f(M)$ is decomposable, e.g., \texttt{SUM}, \texttt{AVG}, \texttt{Variance}, we can derive $ts(R-\sigma_{\scaleto{E}{3.5pt}}R)$ by using $ts(R)$ and $ts(\sigma_{\scaleto{E}{3.5pt}}R)$. Since \sys is designed to integrate with existing interactive data analysis tools like PowerBI, data cube is typically maintained in memory and thus we can easily access $ts(R)$ and $ts(\sigma_{\scaleto{E}{3.5pt}}R)$ from the data cube.

                {\em Time complexity:} for each segment $[p_{j'}, p_j]$ and explanation $E$, it takes $O(1)$ for computing the difference score $\gamma(E)$. There is in total $n^2$ segments and $\epsilon$ explanations, thus the total time complexity for module (a) is $O(\epsilon \cdot n^2)$.

            \stitle{The Cascading Analysts (CA) Algorithm.}
            Module (b) is responsible for calculating top-explanation $\E_m^*$ for each segment $[p_{j'}, p_j]$.
            \sys employs \cite{Ruhl2018TheCA} to identify top-explanation (Definition~\ref{def:top-m}).
            In a nutshell, the CA algorithm simulates what a data analyst would perform during data analysis: recursively perform drill-down operations in dimensions and select data slices they are interested in.
            Each data slice is summarized by some conjunction predicate, i.e., an explanation in our context. Different from manual data analytics, here the drill-down dimensions and data slices are selected via dynamic programming to maximize the total diff score $\gamma(E)$ under the constraint that the number of data slices is $\leq m$.

            \begin{figure}[h]
                \centering
                \includegraphics[width=\columnwidth]{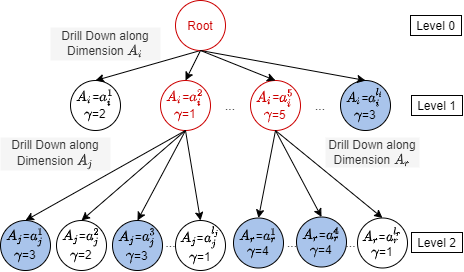}
                \caption{Illustration of Cascading Analyst}
                \label{fig:ca}
            \end{figure}

            Given three explain-by attributes $\A$=$\{A_i, A_j, A_r\}$, Figure~\ref{fig:ca} illustrates the CA algorithm for identifying top-5 explanations.
            Each node in Figure~\ref{fig:ca} corresponds to an explanation $E$ and is associated with its diff score $\gamma(E)$ obtained from module (a).
            For instance, the left-most node at level one denotes explanation $E=(A_i$=$a_i^1)$ and the left-most node at level two denotes explanation $E=(A_i$=$a_i^2\&A_j$=$a_j^1)$. To identify top-m non-overlapping explanations, the algorithm starts from the root node with $m$ quotas. It enumerates the first drill-down dimension, e.g., dimension $A_i$ as shown in Figure~\ref{fig:ca}, as well as the quota assigned to each drill-down sub-tree, e.g., two out of five is assigned to the subtree rooted at node $(A_i$=$a_i^2)$ and $(A_i$=$a_i^5)$ respectively and another one is assigned to the subtree rooted at node $(A_i$=$a_i^{l_i})$.
            Again, for the sub-tree rooted at node $(A_i$=$a_i^2)$ with two quotas, the algorithm enumerates next dimension to drill down (i.e., $A_j$ in Figure~\ref{fig:ca}), and assigns quota to each drill-down sub-tree (i.e., one to $(A_i$=$a_i^2\&A_j$=$a_j^1)$ and one to $(A_i$=$a_i^2\&A_j$=$a_j^3)$). This process is conducted recursively. The enumeration of drill-down dimension and quota assignment are performed via dynamic programming to maximize the total score $\sum_{\scaleto{E}{4pt}} \gamma(E)$ under the constraint that the total quota is $\leq m$.  In Figure~\ref{fig:ca}, the algorithm returns top-5 explanations (nodes in blue) with maximum $\sum_{\scaleto{E}{4pt}} \gamma(E)$=3+3+4+4+3=$17$. Please refer to \cite{Ruhl2018TheCA}~\footnote{\cite{Ruhl2018TheCA} also works with the alternative Definition~\ref{def:top-m} in footnote 1.} for details.

                {\em Time complexity:} the CA algorithm~\cite{Ruhl2018TheCA} takes $O(\epsilon\cdot |\A|\cdot m^2)$ per segment, where $\epsilon$ is the total number of candidate explanations,
        $|\A|$ is the number of explain-by attributes, and $m$ is a user-specified explanation number. By default, $m$ is set as 3. Since there are in total $n^2$ segments, the total time complexity is $O(\epsilon \cdot |\A| \cdot m^2 \cdot n^2)$

            \stitle{K-Segmentation.}
            Module (c) is responsible for identifying the best K-segmentation scheme $\P^*_K$ that minimizes the total variance. As described in Section~\ref{ssec:design-var} (Eq.~\ref{eqn:distance}), we first compute $dist(P_k, P_x)$ based on the top-explanations $\E_m^*$ on centroid segment $P_k$=$[p_{j'},p_j]$ and object $P_x$=$[p_x,p_{x+1}]$ obtained from module (b); next, $\var(P_k)$ is calculated for every segment based on Eq.~\ref{eqn:variance}; lastly, DP is utilized for deriving $D(n, K)$ and the optimal segmentation scheme $\P_K^*$.

                {\em Time complexity:} for each pair of $(P_k, P_x)$, calculating $dist(P_k, P_x)$ takes $O(m)$ in looking at top-$m$ explanations. There are $(n$-$l)$ centroid segments $P_k$ of length \footnote{A segment $[p_i, p_j]$ has length $(j-i)$.} $l$ where $1$$\le$ $l$<$n$ and each $P_k$ of length $l$ contains $l$ objects $P_x$. In total, we have $\sum_{l=1}^{n-1} l(n-l)$=$O(n^3)$ pairs of $(P_k, P_x)$ and thus the time complexity for calculating all distance is $O(m\cdot n^3)$. Similarly, the total time complexity for calculating variance $\var(P_k)$ of all $P_k$ is $O(n^3)$. With $\var(P_k)$ available, DP takes $O(n\cdot nK)=O(n^2K)$, as each step in Eq.~\ref{eq:k_segmentation_dp} involves $O(n)$ for enumerating the last cut's position and there is in total $nK$ steps. Thus, the complexity of module (c) is $O(m\cdot n^3 + K\cdot n^2)$.

        \stitle{Takeaway.}
        To sum up, the total time complexity is $O(\epsilon\cdot n^2$+$\epsilon\cdot |\A|\cdot m^2 \cdot n^2$+$m\cdot n^3$+$K\cdot n^2)$ --- it grows linearly to $\epsilon$ and $|\A|$, quadratic to $m$, and cubed to $n$. In general, $m$, $K$, and $|\A|$ are small due to user's limited perception.
        Thus, the runtime depends mostly on the number of candidate explanations $\epsilon$ and the size of the aggregated time series $n$. Next, we focus on reducing $\epsilon$ and $n$ for speedup.

\input{content/optimizations.tex}

%% file: content/optimizations.tex
\subsection{Optimizations} \label{ssec:opt}
As the takeaway above describes, the runtime largely depends on the number of candidate explanations $\epsilon$ and the time series length $n$. For \texttt{Liquor} dataset used in Section~\ref{sec:exp}, $\epsilon$ is around 5000 even when only two explain-by attributes are considered, i.e., $|\A|$=2, and $n$ is around 300. Next, we introduce two optimizations for speedup: {\em (I.)} \texttt{guess-and-verify} to reduce $\epsilon$ and {\em (II.)} \texttt{sketching} to reduce $n$.

\subsubsection{\texttt{Guess-and-verify}}
As discussed in Section~\ref{ssec:pipeline}, the CA algorithm is one of the bottlenecks in our pipeline. \texttt{Guess-and-verify} is designed to reduce the input explanation number ($\epsilon$) in CA.

\stitle{High-level Idea.}
To ensure non-overlapping, CA algorithm recursively drills down dimensions and selects data slices (explanations) as shown in Figure~\ref{fig:ca}.
The intuition behind \texttt{guess-and-verify} is that explanation $E$ with higher diff score $\gamma(E)$ is more likely to be in the top-$m$ {non-overlapping} explanations $\E_m^*$. Thus, instead of using all candidate explanations as the input of CA algorithm, we can take a {\em guess} and limit the input to only include top explanations with the highest diff score. A smaller input size can dramatically reduce the runtime of the CA algorithm, but the returned results $\overline{\E_m^*}$ may not be the optimal top-$m$ non-overlapping explanations $\E_m^*$. To mitigate this, we {\em verify} whether the returned result is optimal. This process is repeated until the result is verified to be optimal.

\input{content/guessverifydetail.tex}

\subsubsection{\texttt{Sketching}}
As discussed in Section~\ref{ssec:pipeline}, the time complexity in each module is at least quadratic to the time series size $n$. This is because each point is treated as a candidate cutting position in K-Segmentation and thus the total number of segments involved in each module is $O(n^2)$. Hence, reducing the number of candidate cutting positions can dramatically improve the efficiency. \texttt{Sketching} is designed for this purpose, as depicted in Figure~\ref{fig:pipeline}.

\stitle{High-level Idea.}
Given a time series with $n$ points, K-Segmentation aims to select $(K$-$1)$ cutting points out of $(n$-$2)$ non-endpoints. Our intuition is that some points are worse-suitable to be used as the cutting points and can be eliminated in a more cost-effective manner; next, since the remaining points is of a much smaller size, it is affordable to input them in our solution pipeline (Section~\ref{ssec:pipeline}). We call the remaining points \texttt{sketch}, as its role is to represent the original $n$ points in the given time series. In particular, \texttt{Sketching} consists of two phases: {\em (I.)} \texttt{sketch} selection and {\em (II.)} pipeline with \texttt{sketch}.
We propose to select sketch using our proposed pipeline in Section~\ref{ssec:pipeline}, but with the constraint that each segment's length to be within $L$, where $L << N$.

\input{content/sketchdetail.tex}

%% file: content/guessverifydetail.tex
Specifically, we first sort all explanations in descending order of $\gamma(E)$, denoted as $\chi$. Next, we go through the following two phases iteratively: {\em (1) guess} and {\em (2) verify}. In each iteration, first take a guess on the input size $\bar{m}$; then run the CA algorithm with the first $\bar{m}$ explanations in $\chi$ and obtain the result $\overline{\E}$; verify if $\overline{\E}$ is optimal. 

\begin{figure}[h]
    \centering
    \includegraphics[width=\columnwidth]{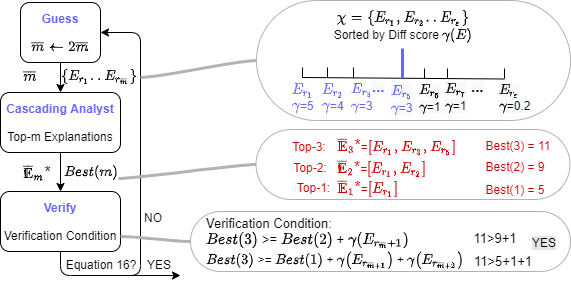}
    \caption{Guess-and-Verify}
    \label{fig:guess-and-verify} 
\end{figure}

\stitle{Guess.} 
Naively selecting the first $m$ explanations in $\chi$ might result in explanations that overlap with each other, though the total difference score is the largest. Alternatively, we hypothesize that the answer of the CA algorithm comes from the top $\bar{m}$ candidate explanations in $\chi$. If the returned result passes the verification condition, terminate; otherwise, increase $\bar{m}$ to $2\bar{m}$ as shown in Figure~\ref{fig:guess-and-verify}. 

\stitle{Verify.} 
Taking the first $\bar{m}$ explanations in $\chi$ as the input, the result $\overline{\E_m^*}$ returned by CA algorithm is not guaranteed to be optimal since some explanation $E\in\E_m^*$ might rank after $\bar{m}$ in $\chi$. To ensure the optimality, we design a sufficient condition such that once this condition is satisfied, it is guaranteed that the returned result is optimal. Let $\chi$=$[E_{r_1}, E_{r_2}..,E_{r_{\epsilon}}]$ be the ordered explanation list ranked by $\gamma(E)$. With $[E_{r_1}, E_{r_2}..,E_{r_{\bar{m}}}]$ as input, let $\overline{\E}^*_m$ be the top-$m$ non-overlapping explanations returned by CA algorithm and $Best[m]$ be the corresponding total difference score. Given $m$, the CA algorithm not only returns $Best[m]$, but also $Best[m']$ for every 1$\le$ $m'$<$m$ as a side product of dynamic programming. $Best[0]=0$. With these notations, we can now present the verification condition in Eq.~\ref{eq:verification}. The high-level idea is that each explanation can be categorized into the following two classes: (1) with rank $\leq$ $\bar{m}$, and (2) with rank >$\bar{m}$. Thus, we can upper bound the total difference score of each candidate {\em $m$ non-overlapping explanations} by the right-hand side of Eq.~\ref{eq:verification}, where the first term corresponds to the score of class (1) and the second term upper bounds the score of class (2). Hence, if we ensure that the current best solution (left-hand side of Eq.~\ref{eq:verification}) has higher total score than all candidate solutions (right-hand side of Equation~\ref{eq:verification}), we can safely terminate and output the optimal top-explanation. Proofs are omitted due to space limitation.

\begin{equation} \label{eq:verification}
    Best[m] \ge Best[m'] + \sum_{1 \le j \le m-m'} \gamma(E_{r_{\bar{m} + j}}) \texttt{   } \forall  0\le m' < m
\end{equation}

{\em Time complexity:} {\bf \texttt{guess-and-verify} decreases the complexity from $O(\epsilon \cdot|\A|\cdot m^2 \cdot n^2)$ to $O(\bar{m} \cdot|\A|\cdot m^2 \cdot n^2)$ at the best case}. Empirically, when $m=3$, we initialize $\bar{m}=30$.

%% file: content/sketchdetail.tex
\stitle{\texttt{Sketch} Selection.}
There are two main requirements for \texttt{sketch} selection. First, the process should be efficient; second, the selected \texttt{sketch} should contain promising points for small-variance K-segmentation scheme (Eq.~\ref{eqn:variance}).
Strawman approaches like random sampling or evenly spaced sampling are fast, but does not meet requirement two. To satisfy both, we propose to utilize our proposed pipeline in Section~\ref{ssec:pipeline}, but with some constraint. As we will detail below, the constraint is used to reduce the pipeline runtime (requirement one); and the pipeline is used to identify promising cutting points for minimizing the total variance in Eq.~\ref{eqn:variance} (requirement two).

As discussed, the number of segments considered in our solution pipeline (Section~\ref{ssec:pipeline}) is $O(n^2)$. To alleviate this bottleneck, we can restrict each segment's length to be within $L$, where $L << N$. In this way, we only need to compute the diff score $\gamma(E)$ (module a), top-explanations $\E_m^*$ (module b), distance $dist(P_k, P_x)$ and variance $\var{P_k}$ (module c) for segments with length $\leq L$. This reduces the total number of considered segments from $O(n^2)$ to $O(Ln)$. More specifically, to derive a \texttt{sketch} of size $|S|$, we set $K=|S|$ in our K-segmentation pipeline and set the maximum segment length threshold as $L$. Empirically, we set $L$=$min(0.05N,20)$ and $|S|$=$\frac{3n}{L}$.

{\em Time complexity:} In phase I (\texttt{sketch} selection), {\em module (a):} the time of computing the diff scores turns to $O(\epsilon\cdot L\cdot n)$; {\em module (b):} the time of CA algorithm turns to $O(\epsilon\cdot|\A|\cdot m^2 \cdot Ln)$; {\em module (c):} the time of computing distance and variance turns to $O(m\cdot L^2\cdot n)$, and the time of DP turns to $O(L\cdot nK)$. {\bf The total time complexity is reduced by at least $\frac{L}{n}$, compared to the pipeline without constraint.}
In phase II (pipeline with \texttt{Sketch}), {\em module (a):} the time of computing the diff scores turns to $O((\epsilon\cdot |S|^2)$; {\em module (b):} the time of cascading analyst turns to $O(\epsilon\cdot|\A|\cdot m^2 \cdot|S|^2)$; {\em Module (c):} the time of computing distance and variance scores turns to $O(m\cdot|S|^2\cdot n)$, and the time of DP turns to $O(K\cdot |S|^2)$. {\bf The total time complexity is reduced by $(\frac{|S|}{n})^2$, compared to the pipeline without \texttt{sketch}.}

%% file: content/extension.tex
\section{Optimal Selection of $K$}\label{sec:optimal-k}

In real-world datasets, it is hard for users to specify the number of segments $K$ in advance.
By varying segment number $K$, \sys outputs segmentation schemes with different variance scores, generating a $K$-$Variance$ curve. The left-hand side of Figure~\ref{fig:covid-total} illustrates an example. Intuitively, $K$-$Variance$ curves decrease monotonically as the increase of $K$. At an extreme, when $K$=$n$-$1$, the total variance reaches a minimum score of zero. Furthermore, the total variance score drops quickly when $K$ is small and slows down when $K$ grows larger. This indicates that the marginal improvement of increasing $K$ becomes smaller when $K$ is large. Also, a larger $K$ brings about too many segments, which would exceed user perception limitations. Thus, our goal is to identify the optimal $K$ with relatively low variance and keep the segmentation scheme concise.

Such a task is well-studied in the machine learning community~\cite{kodinariya2013review}.
We borrow the idea of a well-known method named the ``Elbow method''~\cite{ng2012clustering, kodinariya2013review} which picks the ``elbow point'' of the $K$-$Variance$ curve as the optimal $K$.
We use a task-agnostic algorithm~\cite{satopaa2011finding} to automatically determine the ``elbow point'' of our $K$-$Variance$ curve. This algorithm first normalizes the curve to be from $(0,0)$ to $(1,1)$.
Then, it picks {\small $K^*$=$\argmax_{K}[\totalvar(K)-K]$} as the ``elbow point" where {\small$\totalvar(K)$} denotes the normalized total variance when the segment number is $K$.

In our implementation, we collect the dynamic programming results $D(n, K)$ varying $K$ from 1 to 20, plot the $K$-$Variance$ curve, and then choose the elbow point. We note that compared to calculating $D(n, K=20)$, collecting $D(n, K)$ with varying $K$ from 1 to 20 does not add extra cost, since $D(n, K)$ gets generated for $1\leq K <20$ during the dynamic programming process of $D(n, K=20)$.
We constrain $K$ to be at most 20 due to user perception limitation: when $K$ is too large, e.g., $K\geq 20$, it would be hard for users to interpret the explanation results.
We admit that when the time series is long, i.e., $n$ is large, restricting $K$ under 20 might not return the explanations at the finest granularity, but we argue that explanations at coarse grain with $K \le 20$ is a better choice considering user perception limit.
Empirically, we observe that \sys chooses 6 or 7 segments in most cases in our real-world experiments (\Cref{ssec:expreal}).

%% file: content/exp.tex
\section{Experiments} \label{sec:exp}
In this section, we answer two questions: (1) how effective is \sys in identifying the evolving explanations; (Section~\ref{exp:synthetic} and \ref{ssec:expreal}) (2) how fast is \sys (Section~\ref{ssec:exp_efficiency}).

The current \sys is implemented in C++. All experiments below are
run single-threaded on a Macbook Air 2020 with Apple M1 chip 8‑core CPU and 16GB memory.

\subsection{Datasets}
We introduce the datasets used in the experiments. 
\subsubsection{Synthetic datasets}
We synthesize 20 different datasets with 7 different levels of $SNR_{dB}$ (\cref{sec:syn}). In total, we have 140 different datasets. 
The aggregated time series is {\small {\tt SELECT $\T$, count(sales) FROM R GROUP BY $\T$}} and the explain-by attribute is {\tt category}.
\subsubsection{Real-world datasets} 
We use the three real-word datasets in the motivating examples of \Cref{sec:intro}. Here, we briefly go through the aggregated time series, explain-by attributes. In practice, we expect users to provide explain-by attributes based on their domain knowledge.

\paragraph{\textbf{Covid}} \cite{dong2020interactive} records the daily/total confirmed cases of 58 states. Naturally, there are two aggregated time series: \textcircled{1} the covid total confirmed cases trend, {\small{\tt SELECT date, SUM(total-confirmed-cases) FROM Covid GROUP BY date}}; \textcircled{2} the covid daily confirmed cases trend, {\small{\tt SELECT date, SUM(daily-confirmed-cases) FROM Covid GROUP BY date}}. We choose {\tt state} as our explain-by attribute to answer "which {\tt states} are the main contributors to the rises or drops?".

\paragraph{\textbf{S\&P 500}} contains 503 company\footnote{S\&P 500 has 505 components and the components are adjusted along the time. We select the 503 companies that are in the component list during the whole period.} stock price ({\tt price}) and free-float shares ({\tt share}) from 2020-1-1 to 2020-10-1. Based on the S\&P 500 index formula\cite{sp500calc}, we derive the S\&P 500 index's time serie as 
${\small{\tt SELECT~ date,~ \frac{SUM(price*share)}{divisor} ~ AS}}$ {\small{\tt SP500-index ~ FROM ~Sp500}}
 {\small{\tt GROUP~ BY}} {\small{\tt date}}, where divisor is a constant. We try to explain the S\&P 500 index's crashes and rebound using the hierarchical explain-by features - {\tt category, subcatagory, stock}.

\paragraph{\textbf{Liquor}} contains liquor purchase transactions in Iowa from 2020-1-2 to 2020-6-30. The time series is
{\small{\tt SELECT date, SUM(Bottles Sold) FROM Liquor GROUP BY date}}.
We pick four attributes out of 24 attributes as our explain by features:  {\tt Bottle Volume (ml)} -- the size of each bottle in a purchase (e.g., 750ml); {\tt Pack} -- the number of bottles per pack (e.g., 6); {\tt Category Name} -- the category of the purchased liquor (e.g., American Flavored Vodka); {\tt Vendor Name} -- the vendor of the purchased liquor (e.g. Phillips Beverage). Below, we use BV, P, CN, and VN to represent them respectively for short.

\subsection{Baseline}\label{sec:baseline}

\sys is the first explanation-aware segmentation to surface the evolving explanations for time series. The closest segmentation works are the bottom up algorithm ({\it Bottom-Up}) which performs best overall in piecewise linear approximation~\cite{keogh2004segmenting} and recent semantic segmentation algorithm - {\it FLUSS}~\cite{gharghabi2017matrix} and {\it NNSegment}~\cite{sivill2022limesegment}. All these methods are explanation-agnostic, partition time series solely based on the visual shapes, and require segment number as input. For a fair comparison, the $K$ for baselines is either given or borrowed from \sys's results. Implementation wise, we reproduce {\it Bottom-Up}  based on the pseudo-code in Keogh et al~\cite{keogh2004segmenting}. {\it FLUSS} is implemented using Stump library~\cite{law2019stumpy} and {\it NNSegment} is implemented using the authors' code~\cite{sivill2022limesegment}.
\input{content/expsyn}

\input{content/expreal}
\input{content/expperf}

%% file: content/expsyn.tex
\subsection{Explanations of Synthetic Datasets} \label{exp:synthetic}
We evaluate \sys's effectiveness on synthetic datasets and perform quantitative comparisons between \sys and the three baselines.
For fair comparison, we adopt the oracle segment number $K$ of the ground truth, and we run \sys and baselines with known $K$.

\stitle{Metric} We propose a metric to compare the effectiveness of these methods by quantifying the distance between these methods' output and ground truth. We calculate the edit distance between outputs and ground truth. Since different datasets have different segment number $K$ and time series lengths $n$, we normalize our edit distance by $K$ and $n$. The lower the metric is, the more effective the method is. We term this metric $distance~percent(\%)$.

\begin{figure}[h]
    \footnotesize
    \centering
    \includegraphics[width=\linewidth]{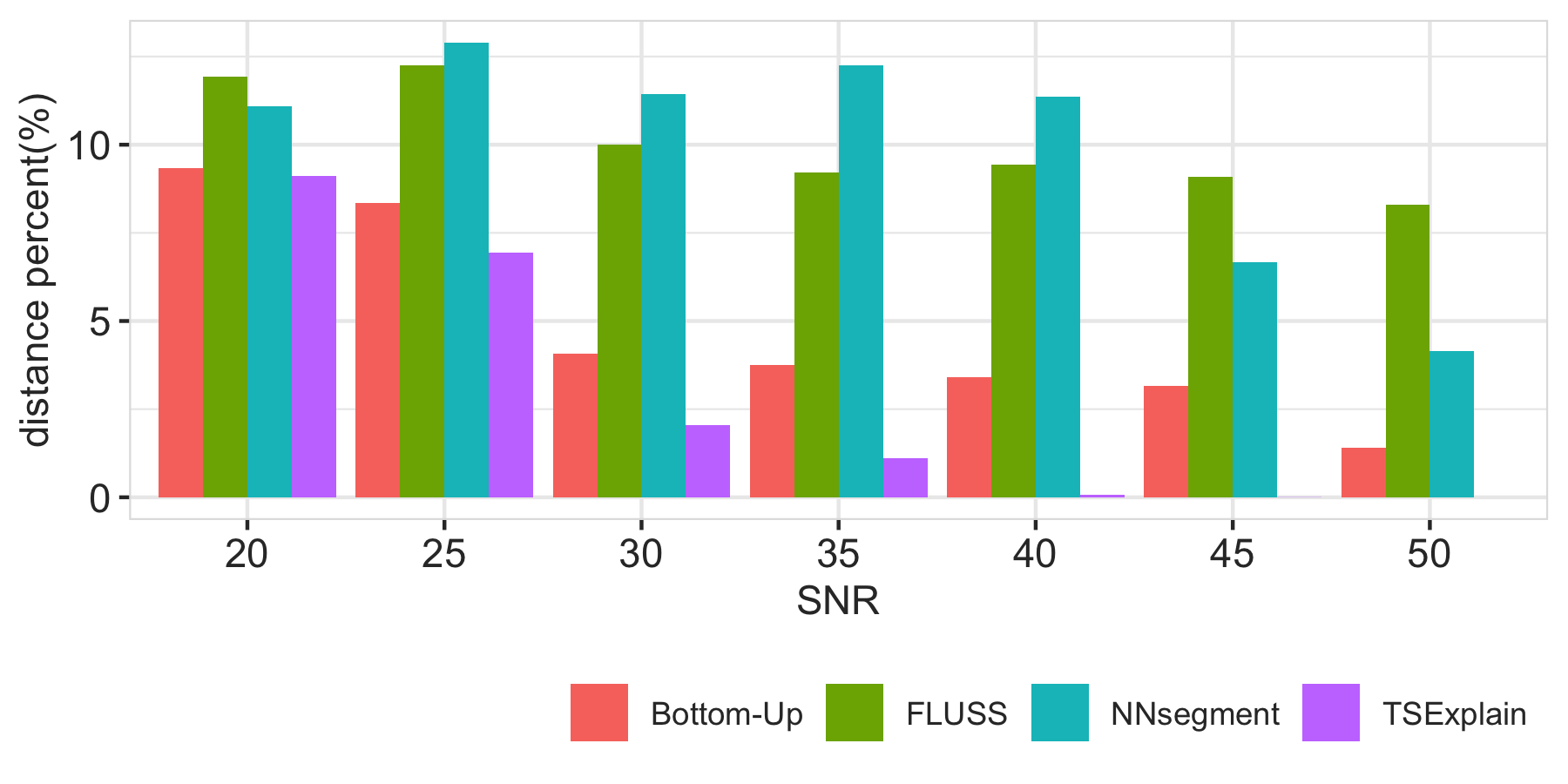}
    \vspace{-3mm}
    \caption{Distance percentage of \sys and baselines}
    \label{fig:dis}
    \vspace{-4mm}
\end{figure}

\Cref{fig:dis} shows the comparison between \sys and baselines. As {\it FLUSS} and {\it NNSegment} both involves parameter, i.e. period and window size, we try mutiple parameters and report the best overall results we found.
The x-axis is SNR, and the y-axis is the $distance~percent(\%)$. We report the average distance percent for each SNR level. As we can see, \sys always has the best performance than all three baselines and {\it Bottom-Up} is the most comparable baseline among all three. When SNR $\textgreater$ 35, the $distance~percent(\%)$ of \sys is close to 0, indicating for cleaner datasets, \sys's output is almost the same as ground truth. However, the baselines are incapable to detect ground truth even with clean datasets.

\stitle{Takeaway.}
For synthetic datasets, \sys performs much more effectively than all baselines. {\it Bottom-Up} is the most comparable baseline. For less noisy dataset, \sys can accurately detect the ground truth, while baselines can not.

%% file: content/expreal.tex
\subsection{Case Study on Real-World Datasets}\label{ssec:expreal}

This subsection demonstrates \sys's effectiveness on three different real-world datasets: Covid, S\&P500, and Liquor, and gives an illustrative comparison between \sys and baselines. For fair comparison, \sys recommends the $K$ and our baseline uses the same segment number $K$. In this experiments, we focus on the top three explanations and set each explanation's order as 3. For very fuzzy datasets, we apply a moving average to smooth it before explaining it. The moving average window can be customized in the interface's panel as shown in the demo \cite{chen2021tsexplain}.

\subsubsection{Covid}
We explain the two time series in Covid dataset, {\tt total-}{\tt confirmed-cases} and {\tt daily-confirmed-cases} separately.

\begin{figure}
    \centering
    \begin{subfigure}{\columnwidth}
        \centering
        \begin{subfigure}{\columnwidth}
            \centering
            \includegraphics[width=\columnwidth]{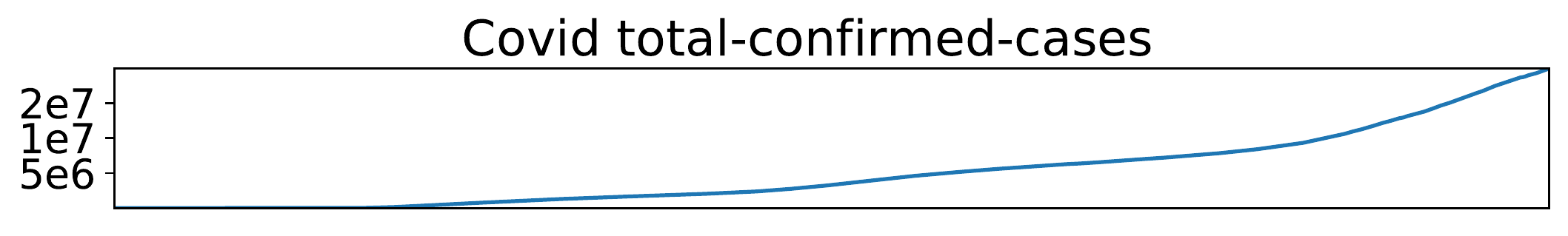}
            \caption{The total confirmed cases trend. }
        \end{subfigure}
        \begin{subfigure}{\columnwidth}
            \centering
            \includegraphics[width=\columnwidth]{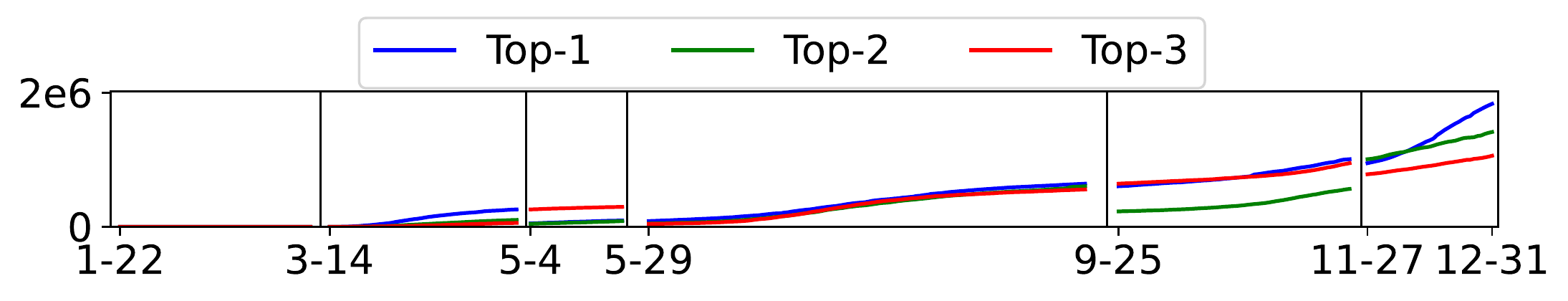}
            \caption{\sys segmentation. }
        \end{subfigure}
        \begin{subfigure}{\columnwidth}
            \centering
            \includegraphics[width=\columnwidth]{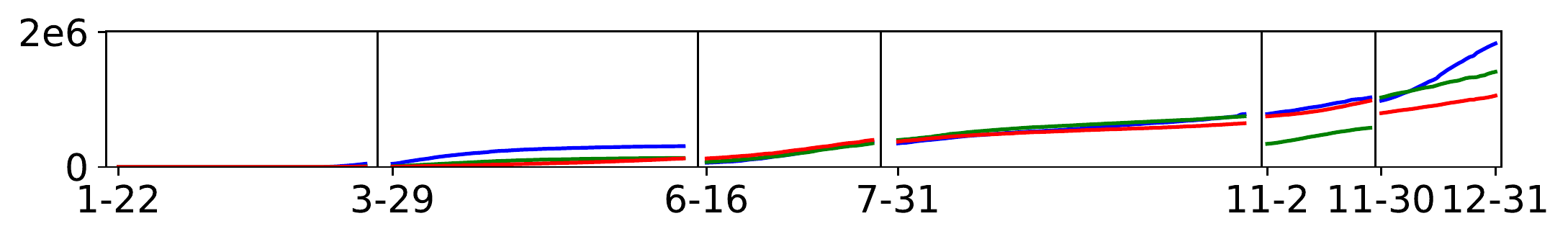}
            \caption{Bottom-Up segmentation. }
            \label{fig: covid-vis}
        \end{subfigure}
        \begin{subfigure}{\columnwidth}
            \centering
            \includegraphics[width=\columnwidth]{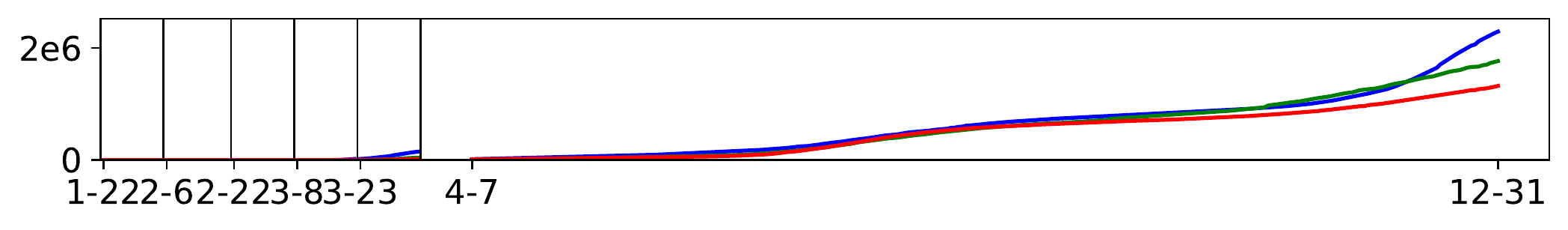}
            \caption{FLUSS segmentation. }
            \label{fig: covid-fluss}
        \end{subfigure}
        \begin{subfigure}{\columnwidth}
            \centering
            \includegraphics[width=\columnwidth]{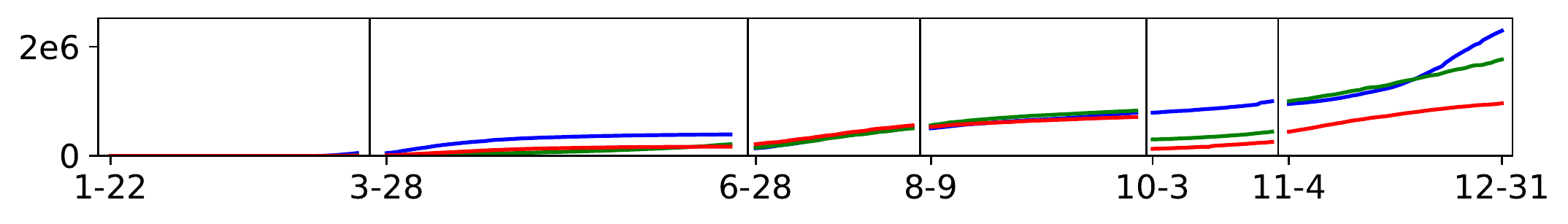}
            \caption{NNSegment segmentation. }
            \label{fig: covid-nnseg}
        \end{subfigure}
    \end{subfigure}
    \caption{Segmentation of total-confirmed-cases. }
    \label{fig:covid-total}
\end{figure}

\paragraph{{\tt Total-confirmed-cases}}
\sys identifies that the optimal $K$ equals 6 based on the optimal selection of K in \Cref{sec:optimal-k}.
\Cref{fig:covid-total} shows \sys's output segmentation scheme with the top three explanations' trend and three baselines' output. Please refer to the legend of \Cref{fig:covid-explain} for the top-3 explanations of \sys.
In \sys's output,  the first segment is from 1/22 to 3/14, where the increase of {\tt total-confirmed-cases}  is due to WA, NY, CA's increase. From 3/14 to 5/4, NY increases the most, followed by NJ and MA. Then, between 5/4 and 5/29, IL, CA, NY slowly increase. Later on, the confirmed cases surge mainly because of the sharp increase in CA, TX, FL, and IL, especially IL increases quickly 

Contrarily, in the baselines, some neighboring segments' explanations are exactly the same, i.e., 6/16--7/31 and 7/31--11/2 in {\it Bottom-Up},  6/28--8/9 and 8/9--10/3 in {\it NNSegment}, as shown in \Cref{fig: covid-vis} and {\Cref{fig: covid-nnseg}}; while {\it FLUSS} segmentation ({\Cref{fig: covid-fluss}}) segments the early time into a lot of small segments which is hard to interpret. In addition, none of the baselines detect the top-explanations' changes from NY during 3/14-5/4 to IL during 5/4-5/29 as reported in news\cite{Illinois}.

\begin{figure}
    \centering
    \begin{subfigure}{\columnwidth}
        \centering
        \begin{subfigure}{\columnwidth}
            \centering
            \includegraphics[width=\columnwidth]{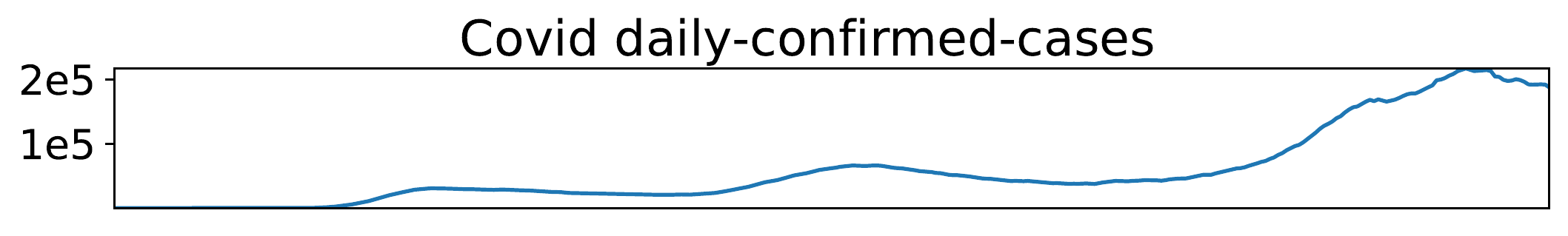}
            \caption{The daily confirmed cases trend. }
        \end{subfigure}
        \begin{subfigure}{\columnwidth}
            \centering
            \includegraphics[width=\columnwidth]{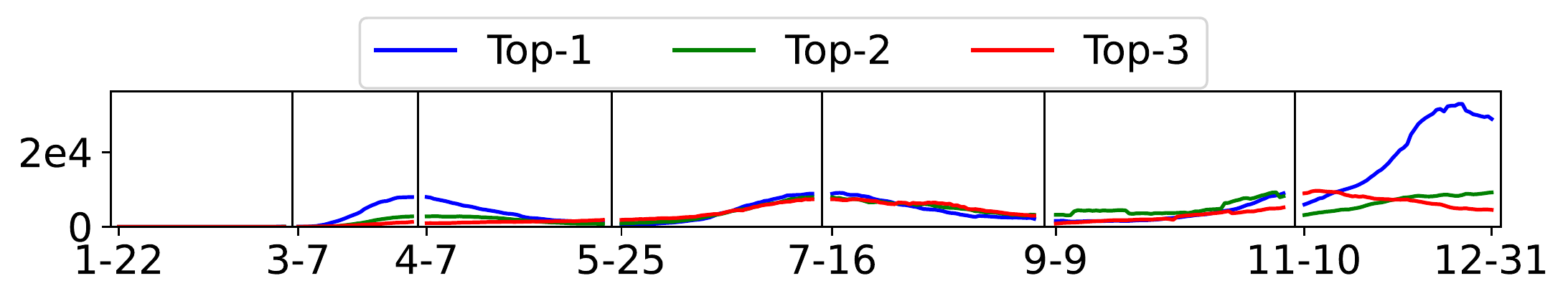}
            \caption{\sys segmentation. }
        \end{subfigure}
        \begin{subfigure}{\columnwidth}
            \centering
            \includegraphics[width=\columnwidth]{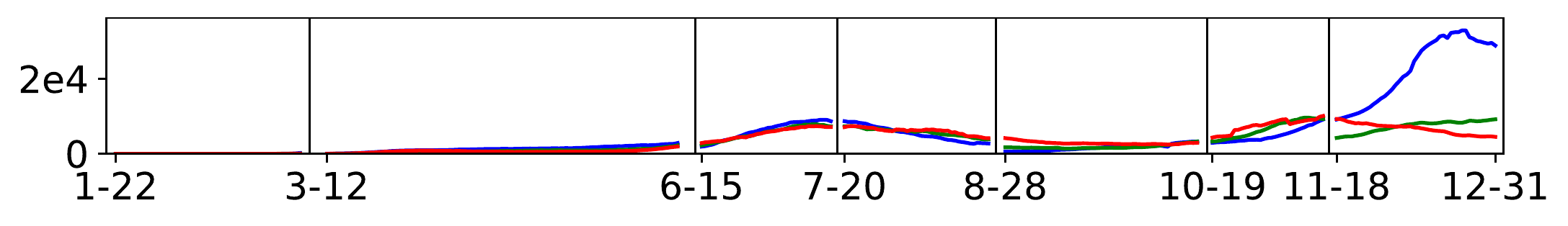}
            \caption{Bottom-Up segmentation. }
        \end{subfigure}
        \begin{subfigure}{\columnwidth}
            \centering
            \includegraphics[width=\columnwidth]{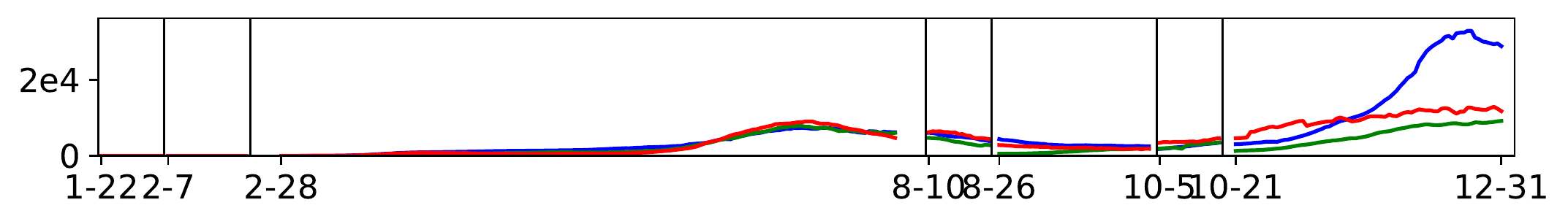}
            \caption{FLUSS segmentation. }
        \end{subfigure}
        \begin{subfigure}{\columnwidth}
            \centering
            \includegraphics[width=\columnwidth]{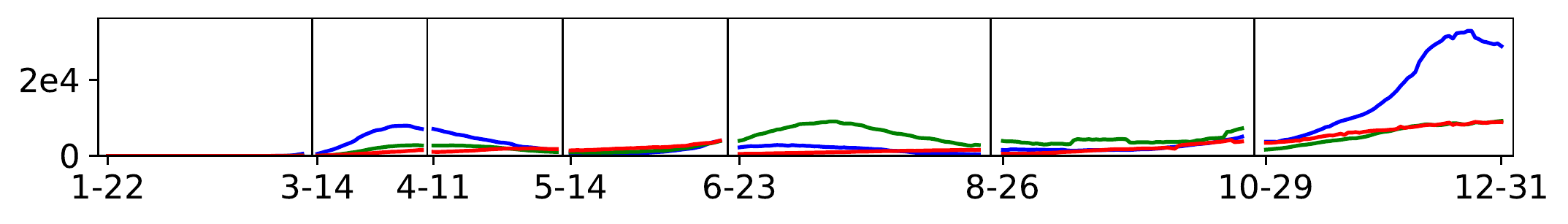}
            \caption{NNSegment segmentation. }
        \end{subfigure}
    \end{subfigure}
    \caption{Segmentation of daily-confirmed-cases.  Left is the K-Variance curve. Right shows aggregated time series(top), \sys results(middle), baseline results(bottom).}
    \label{fig:covid-daily}
\end{figure}
\begin{table}[H]
    \small
    \centering
    \begin{tabular}{lllll}
        \hline
        Segment                     & Top-1 Expl   & Top-2 Expl   & Top-3 Expl      \\ \hline
        1/22 \textasciitilde 3/7    & Washington + & New York +   & California +    \\
        3/7 \textasciitilde 4/7     & New York +   & New Jersey + & Massachusetts + \\
        4/7 \textasciitilde 5/25    & New York -   & New Jersey - & California +    \\
        5/25 \textasciitilde 7/16   & Florida +    & Texas +      & California +    \\
        7/16 \textasciitilde 9/9    & Florida -    & Texas -      & California -    \\
        9/9 \textasciitilde 11/10   & Illinois +   & Texas +      & Wisconsin +     \\
        11/10 \textasciitilde 12/31 & California + & New York +   & Illinois -      \\ \hline
    \end{tabular}
    \caption{Expl. in Fig.~\ref{fig:covid-daily}(middle). +/- denote change effect.}
    \label{table:covid-daily-table}
\end{table}

\paragraph{{\tt Daily-confirmed-cases}} In \Cref{fig:covid-daily} and \cref{table:covid-daily-table}, \sys segments this time series into seven periods.
Specifically, from 3/7 to 4/6, NY, NJ, and MA’s rises contribute to the overall rise in the US. From 4/7 to 5/25, NY and NJ decline dramatically, and \sys captures an interesting pattern that CA starts to rise.
During the holiday season in 2020, namely the last segment, CA and NY surge again while IL declines.   Compared to \sys, the {\it Bottom-Up} segmentation does not detect the changes during 3/7 - 5/25.  The {\it FLUSS} performs very poorly without revealing the explanation during 2/28-9/10 at all.  The {\it NNSegment} segmentation is similar to our \sys explanation. However, it segments the up and down trend during 6/23 -8/26 to a whole segment which makes it users hard to interpret the explanations.

\begin{figure}[h]
    \centering
    \begin{subfigure}{\columnwidth}
        \centering

        \begin{subfigure}{\columnwidth}
            \centering
            \includegraphics[width=\columnwidth]{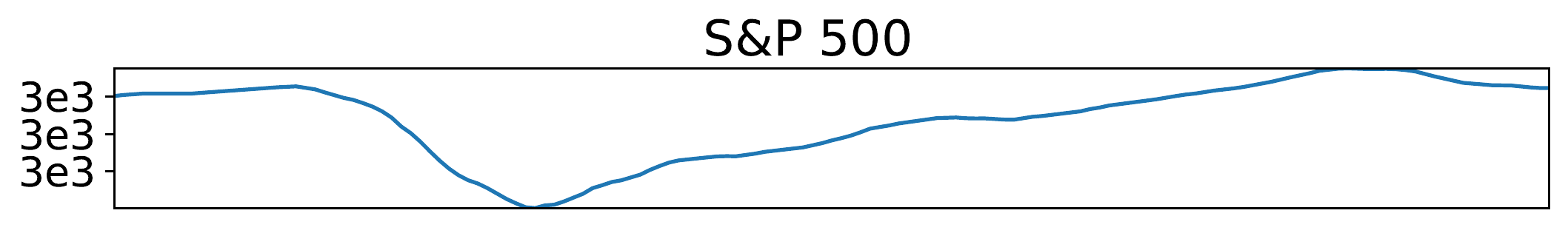}
            \caption{The S\&P500 trend.}
        \end{subfigure}
        \begin{subfigure}{\columnwidth}
            \centering
            \includegraphics[width=\columnwidth]{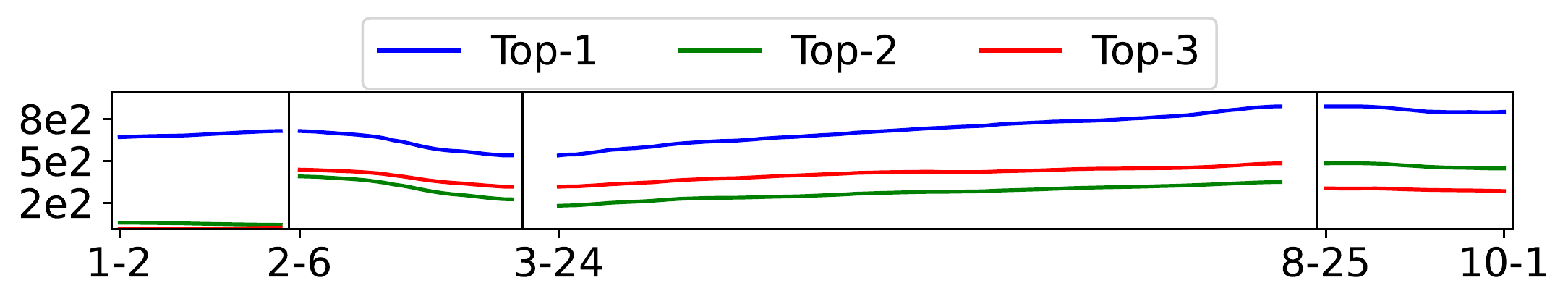}
            \caption{\sys segmentation. }

        \end{subfigure}
        \begin{subfigure}{\columnwidth}
            \centering
            \includegraphics[width=\columnwidth]{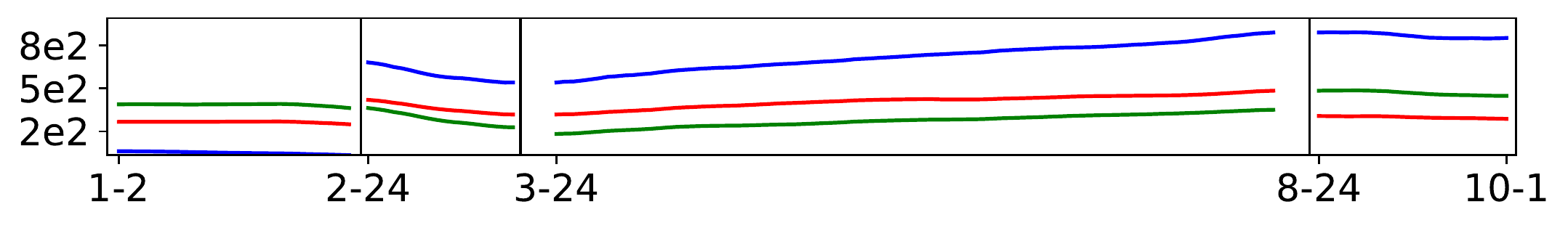}
            \caption{Bottom-Up segmentation. }
        \end{subfigure}
        \begin{subfigure}{\columnwidth}
            \centering
            \includegraphics[width=\columnwidth]{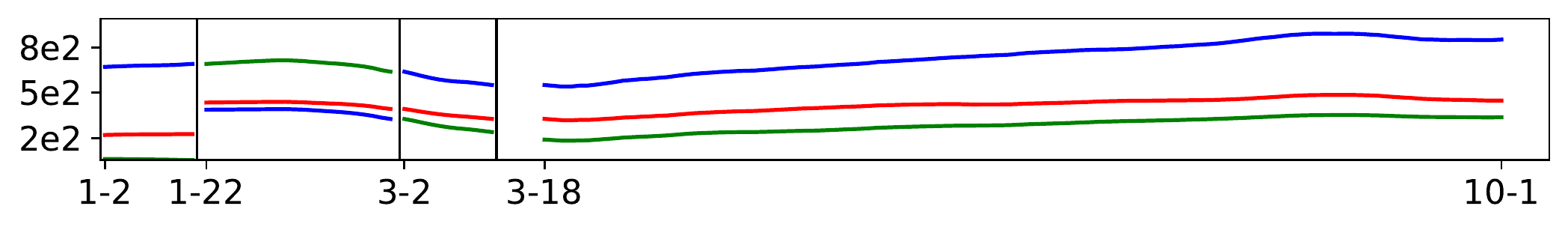}
            \caption{FLUSS segmentation. }
        \end{subfigure}
        \begin{subfigure}{\columnwidth}
            \centering
            \includegraphics[width=\columnwidth]{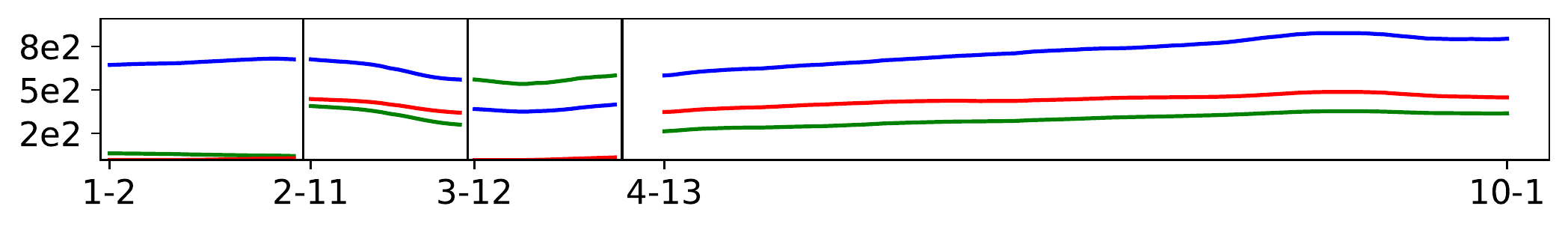}
            \caption{NNSegment segmentation. }
        \end{subfigure}
    \end{subfigure}
    \caption{Segmentation of S\&P 500. Left shows the K-Variance Curve. Right shows aggregated time series(top), \sys results(middle), baseline results(bottom).}
    \label{fig:sp500}
\end{figure}
\begin{table}[h]
    \small
    \begin{tabular}{lllll}
        \hline
        Segment                   & Top-1 Expl   & Top-2 Expl          & Top-3 Expl      \\ \hline    1/2 \textasciitilde 2/6  &  technology + &  energy -                 &  \blue{\textit{internet retail + }}    \\
        2/6 \textasciitilde 3/24  & technology - & financial -         & communication - \\
        3/24 \textasciitilde 8/25 & technology + & consumer cyclical + & communication + \\
        8/25 \textasciitilde 10/1 & technology - & communication -     & financial -     \\ \hline
    \end{tabular}
    \caption{\sys Explanations of S\&P 500 in \Cref{fig:sp500}(middle). All the explanations are related to attribute {\tt category}, except \blue{\textit{internet retail + }} is related to {\tt subcategory}. We omit the attribute name for short.  }
    \label{table:sp500-table}
\end{table}

\subsubsection{S\&P 500}
\Cref{fig:sp500} and \Cref{table:sp500-table} show \sys explanation results. \sys finds the elbow point at four and segments the time series into four small segments. Before the market crash from 1/2 to 2/6, the S\&P 500 rises mainly due to the rises of {\tt category} technology and {\tt subcategory} internet retail; meanwhile, {\tt category} energy slightly drops.
Also, \sys recognizes the market crash at 3/24. \sys explains that stocks belonging to {\tt category} technology, financial, and communication contribute the most to the crash. After that, the {\tt category} technology contributes most to the recovery during 3/24 \textasciitilde 8/25 and the drop during 8/25 \textasciitilde 10/1.
We can also discover an interesting fact that financial {\tt category} drops dramatically from 2/6 to 3/24, but does not bounce back a lot as technology and communication {\tt category} do during the stock market recovery. In the {\it Bottom-Up} approach, although it detects the market crashing point on 3/24, however, the decreasing influencers – technology, financial, communication service categories are detected starting from 2/24 which is much later than \sys.
And {\it FLUSS} and {\it NNSegment} do not detect the crash point well and also can not recognize the drop from Aug to Oct. 

\subsubsection{Liquor}

\begin{figure}[h]
    \centering
    \begin{subfigure}{\columnwidth}
        \centering
        \begin{subfigure}{\columnwidth}
            \centering
            \includegraphics[width=\columnwidth]{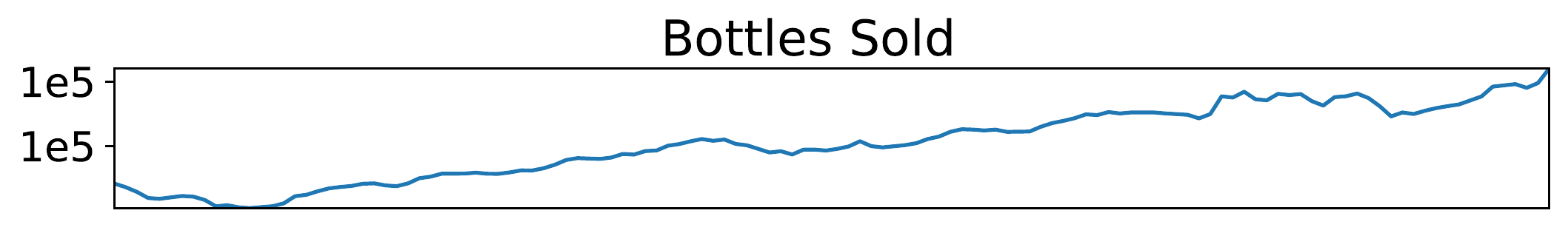}
            \caption{The bottles sold trend.}
        \end{subfigure}
        \begin{subfigure}{\columnwidth}
            \centering
            \includegraphics[width=\columnwidth]{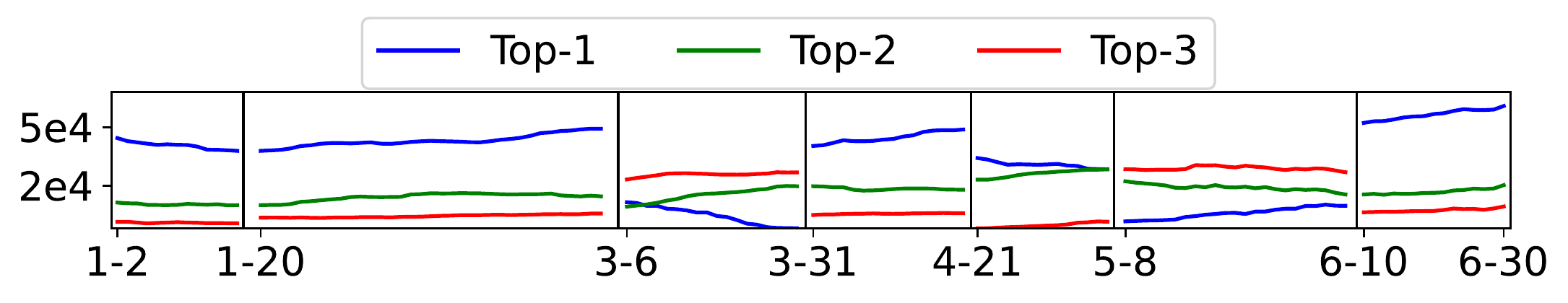}
            \caption{\sys segmentation. }
        \end{subfigure}
        \begin{subfigure}{\columnwidth}
            \centering
            \includegraphics[width=\columnwidth]{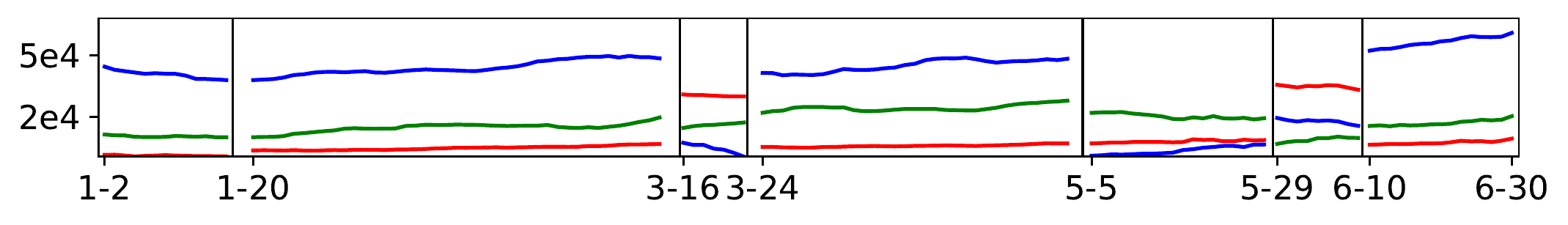}
            \caption{Bottom-Up Segmentation.}
        \end{subfigure}
        \begin{subfigure}{\columnwidth}
            \centering
            \includegraphics[width=\columnwidth]{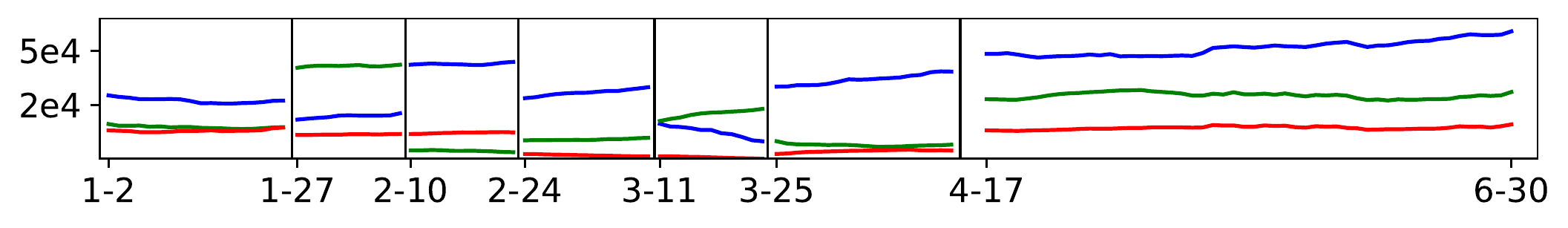}
            \caption{FLUSS segmentation. }
        \end{subfigure}
        \begin{subfigure}{\columnwidth}
            \centering
            \includegraphics[width=\columnwidth]{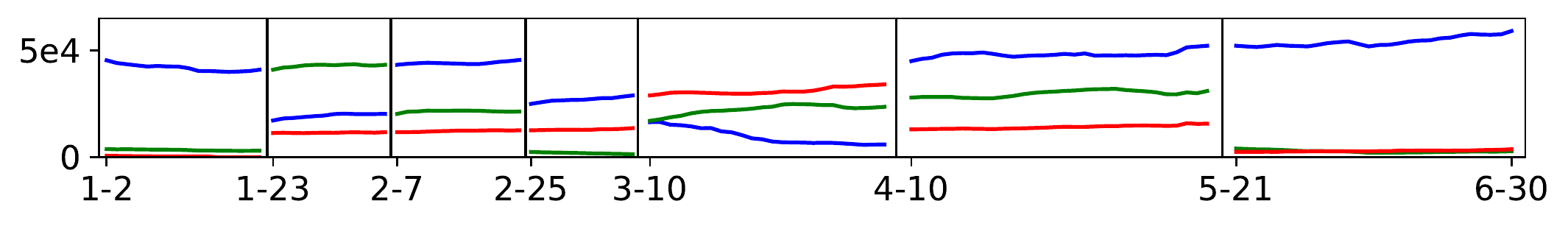}
            \caption{NNSegment segmentation. }
        \end{subfigure}
    \end{subfigure}
    \caption{Segmentation of Liquor. Right shows the K-Variance Curve. Left shows aggregated time series(top), \sys results(middle), baseline results(bottom).}
    \label{fig:liquor}
\end{figure}

    \Cref{fig:liquor} shows that \sys segments the time series into seven segments and \Cref{table:liquor-table} shows each segment's explanations.
    \sys recognizes that overall, from 1/20 to 4/21, the large pack liquor increases a lot, i.e., pack = 12, 24, 48. These indicate that people favor large-pack liquor at the beginning of the pandemic. We also remark that the underlying top explanations can be a combination of up and down trends.
    For example, the main reasons for the overall increase from 3/6 to 3/31  are {\small {\tt BV=1000(-)}}, {\small {\tt BV=1750\&P=6(+)}} and {\small {\tt BV=750\&P=12(+)}}.
    \sys explains this in such way that \textcircled{1} {\small {\tt BV=1000}} decreases sharply, otherwise the overall bottles sold can increase much more; \textcircled{2} {\small {\tt BV=1750\&P=6(+)}} and {\small {\tt BV=750\&P=12(+)}} directly contribute to the increase. Moreover, with some background knowledge, we find that liquor with {\small {\tt BV=1000}} is mainly sold in independent stores. In March, Iowa's close down proclamation\cite{Iowaproclamation} requires restaurants and bars to shut down, and most indepen  dent retailers rely on selling liquor to bars and restaurants. As a result, their business significantly declined. In late April, Iowa Governor issued a proclamation reopening restaurants, and the business of independent liquor stores gradually recovered.  We can see \sys recognizes that from 4/21 to 5/8, {\small {\tt BV=1000ml\&Pack=12}} increases a lot and from 5/8 to 6/10, {\small {\tt BV=1000}} becomes the top increasing explanation indicating that independent stores benefited from the reopening policy. In comparison, the explanations of the {\it Bottom-Up} results look flat, as shown at the bottom of \Cref{fig:liquor}(c), indicating the detected explanations change subtly. This is similar with {\it FLUSS} and {\it NNSegment}. Moreover, the top-2 and top-3 explanations during 3/25-4/17 returned by {\it FLUSS} have only subtle changes in bottles sold. For {\it NNSegment}, the explanations during 1/23-2/7 and 2/7-2/25 are exactly the same, and the changes made by top-2 and top-3 explanations of the last segment are very small as well.
    Also, we remark that although we specify four explain-by attributes, the results are only about BV and P. This indicates that \sys is able to identify interesting attributes and ignore the less interesting ones, i.e. {\small{\tt VN}} and {\small{\tt CN}}.

\begin{table}[h]
    \small
    \centering
    \begin{tabular}{lllll}
        \hline
        Segment                   & Top-1 Expl      & Top-2 Expl     & Top-3 Expl      \\ \hline
        1/2 \textasciitilde 1/20  & P=12 -          & P=6 -          & BV=375\&P=24 -  \\
        1/20 \textasciitilde 3/6  & P=12 +          & P=6 +          & P=48 +          \\
        3/6 \textasciitilde 3/31  & BV=1000 -       & BV=1750\&P=6 + & BV=750\&P=12 +  \\
        3/31 \textasciitilde 4/21 & P=12 +          & BV=1750\&P=6 - & P=24 +          \\
        4/21 \textasciitilde 5/8  & BV=1750\&P=12 - & P=6 +          & BV=1000\&P=12 + \\
        5/8 \textasciitilde 6/10  & BV=1000 +       & BV=1750\&P=6 - & BV=750\&P=12 -  \\
        6/10 \textasciitilde 6/30 & P=12 +          & BV=1750\&P=6 + & P=24 +          \\ \hline
    \end{tabular}
    \caption{\sys Explanations of Liquor in Fig. 12(middle).}
    \label{table:liquor-table}
\end{table}

\subsubsection{Sensitivity to K} \sys adopts the elbow method to detect the optimal $K$.
We observe that a slight change of the optimal $K$ will only bring up a slight shift in the results, e.g., remove or add one cutting point if K minuses/adds 1.

\stitle{Takeaway.}
\sys effectively explains real-world datasets while the baseline defects in that:
\begin{itemize}
    \item Less explanation diversity: the neighboring segments have the same explanations.
    \item Less effectiveness: the underlying key explanations can not be detected, and explanations are detected relatively late.
    \item Less significance: The explanations detected change subtly.
\end{itemize}

%% file: content/expperf.tex
\subsection{ Efficiency Evaluation} \label{ssec:exp_efficiency}
We conduct three experiments to evaluate the efficiency. In the first experiment, we report the breakdown latency and study the impact of our optimization strategies. In the second experiment, we compare the end-to-end runtime between \sys and all three baselines. In the third experiment, we study the scalability of \sys.

\subsubsection{Latency Breakdown and Quality}
\paragraph{\textbf{Methods}}
Besides the two optimizations in \Cref{ssec:opt}, we introduce another straight-forward optimization - {\tt filter}. The filtering protocol works as follows: given an explanation $E$, if each point in its aggregated time series has value smaller than a {\it ratio} of the corresponding value in the overall aggregated time series, we filter this explanation $E$ as its support is low and thus insignificant. We set {\it ratio} as 0.001 by default. 

\begin{figure}
    \centering
    \includegraphics[width=\columnwidth]{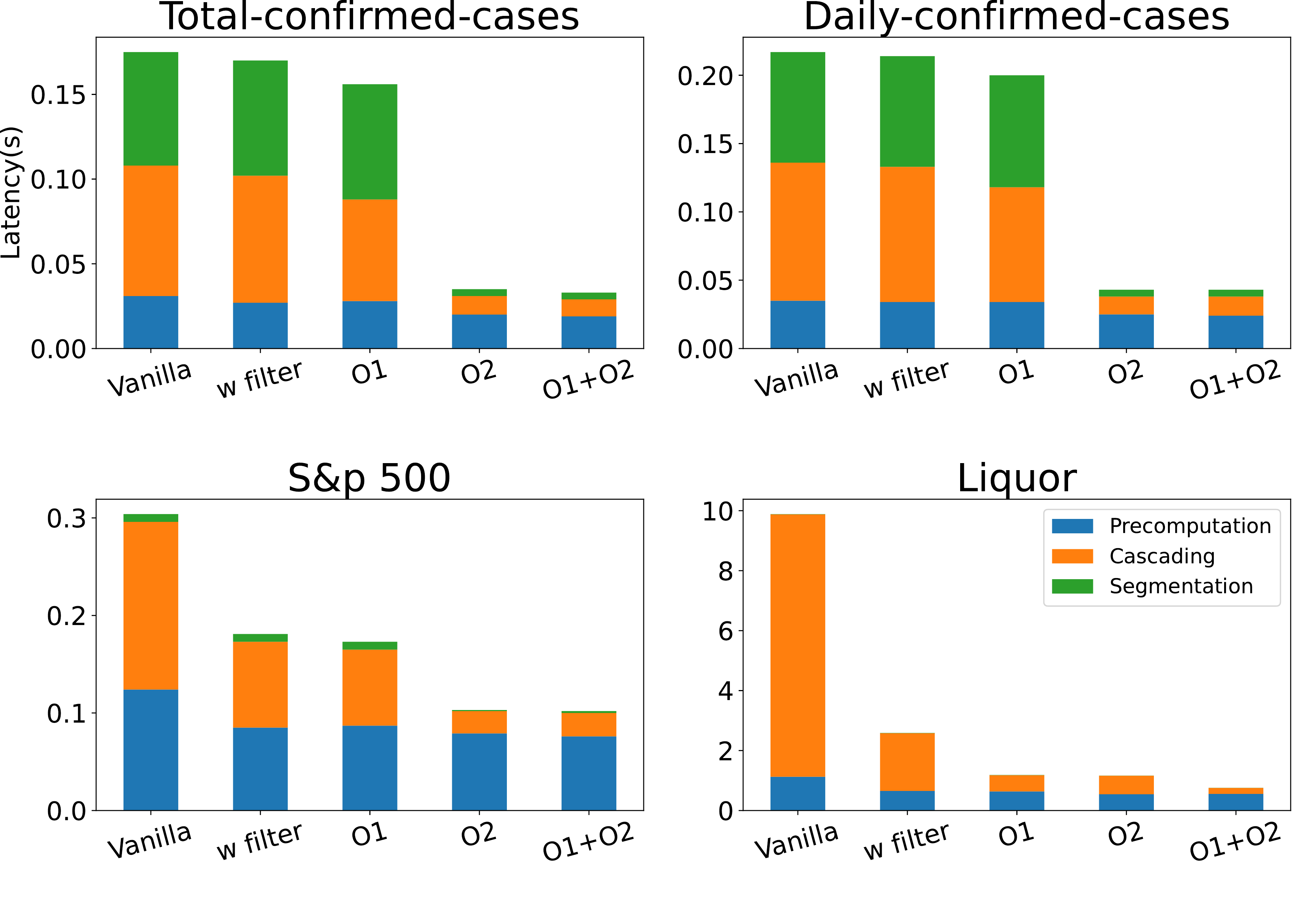}
    \caption{Latency of \sys. }
    \label{fig:performance}
\end{figure}

We study \sys with different optimizations:
\vanSys(short as {\small {\tt Vanilla}}) is the plain version without any optimization;
{\small {\tt w filter}} filters out the explanations below the default {\it ratio};
{\small {\tt O1}} represents the algorithm which applies {\small {\tt guess-and-verify}} after filtering; likewise, {\small {\tt O2}} applies {\small {\tt sketching}} and {\small {\tt O1+O2}} applies both optimizations.
\Cref{table:case-stat} summarizes the statistics of the datasets.
$\epsilon$ is the total candidate explanation number, $n$ is the time series length, and \#record is the dataset record number. We also report the filtered $\epsilon$.
We remark that in this experiment, $K$ is unspecified, and the time of selecting the optimal $K$ via the Elbow Method (\Cref{sec:optimal-k}) is included in the latency (\Cref{fig:performance}).
\begin{table}
\small
    \begin{tabular}{lllll}
    \hline
    dataset        & $\epsilon$ & filtered $\epsilon$ &  n  \\ \hline
  total-confirmed-cases & 58        &54    & 345        \\
    daily-confirmed-cases & 58       &55     & 345         \\
    S\&P 500       & 610         &329   & 151         \\
    Liquor      & 8197       &1812   & 128         \\ \hline
    \end{tabular}
    \caption{Real-world Dataset Statistics.}
    \label{table:case-stat}
\end{table}

\paragraph{\textbf{Latency}} 
\Cref{fig:performance} illustrates the breakdown latency of \sys. 
The overall latency breaks down into three parts -- precomputation(blue), the cascading analysts algorithm(orange), K-segmentation(green) corresponding to three modules in \cref{ssec:pipeline}.

For Covid {\tt total} and {\tt daily-confirmed-cases} dataset, the filtering strategy only slightly improves since the predicate number $\epsilon$ does not change a lot before and after filtering. 
In contrast, the {\tt sketching} in {\tt O2} significantly reduces the latency of the cascading analysts algorithm and K-segmentation. Overall, {\tt O1+O2} reduces the latency of  {\tt total-confirmed-cases} from 175 ms to 33ms, and the latency of  {\tt daily-confirmed-cases} from 217ms to 43ms. 
For S\&P 500 dataset, the filtering strategy reduces the latency to half. {\tt Guess-and-verify} slightly reduces the latency, and together with {\tt sketching} optimization, the overall latency is only 102 ms. 
For the Liquor dataset, the predicate number $\epsilon$ is very large even after filtering. The cascading analyst algorithm is the bottleneck. The {\tt Vanilla} version takes 9.888s. After filtering, it still takes 2.59s. Since the predicate number $\epsilon$ is large, {\tt O1}  ({\tt Guess-and-verify}) takes big effect in this case. Each optimization {\tt O1} or {\tt O2} alone can significantly shrink the runtime to around 1.1s. The two optimizations together reduce the runtime to 756 ms. Although the precomputation is inevitable, the time of the cascading analyst algorithm has been reduced from {\tt vanilla} --- 8.754s to {\tt O1+O2} --- around 200ms. 

\paragraph{\textbf{{Quality}}}
Except {\tt guess-and-verify},  {\tt filter} and {\tt sketching} both approximate the results without formal guarantee. Thus, we study the impact of optimizations in terms of the result quality. 

\Cref{table:sample-quality} illustrates the segmentation scheme's total variance of {\tt O1+O2} compared with the {\tt Vanilla} version. The variances and output segmentation of both algorithms are exactly the same for S\&P 500 and Liquor datasets. For the Covid datasets, the difference is less than 1\% and in the output segmentation, only two cutting points are slightly different --- the corresponding distances are less than four days.  Such a small discrepancy demonstrates that our optimization's effect on result quality is neglectable.

\begin{table}
        \begin{tabular}{llll}
        \cline{1-3}
        dataset         & Variance({\tt Vanilla}) & Variance({\tt O1+O2})  \\ \hline
        total-confirmed-cases  & 22.602      &  22.744             \\
        daily-confirmed-cases  & 91.619       & 91.994             \\
        S\&P 500          & 5.002       & 5.002             \\
        Liquor        & 33.6533        & 33.6533           \\ \hline
        \end{tabular}
        \caption{Quality of optimization strategies.}
        \label{table:sample-quality}
    \end{table}

\subsubsection{End-to-end Efficiency Comparison with Baselines}
\begin{figure}
\footnotesize
\includegraphics[width=\columnwidth]{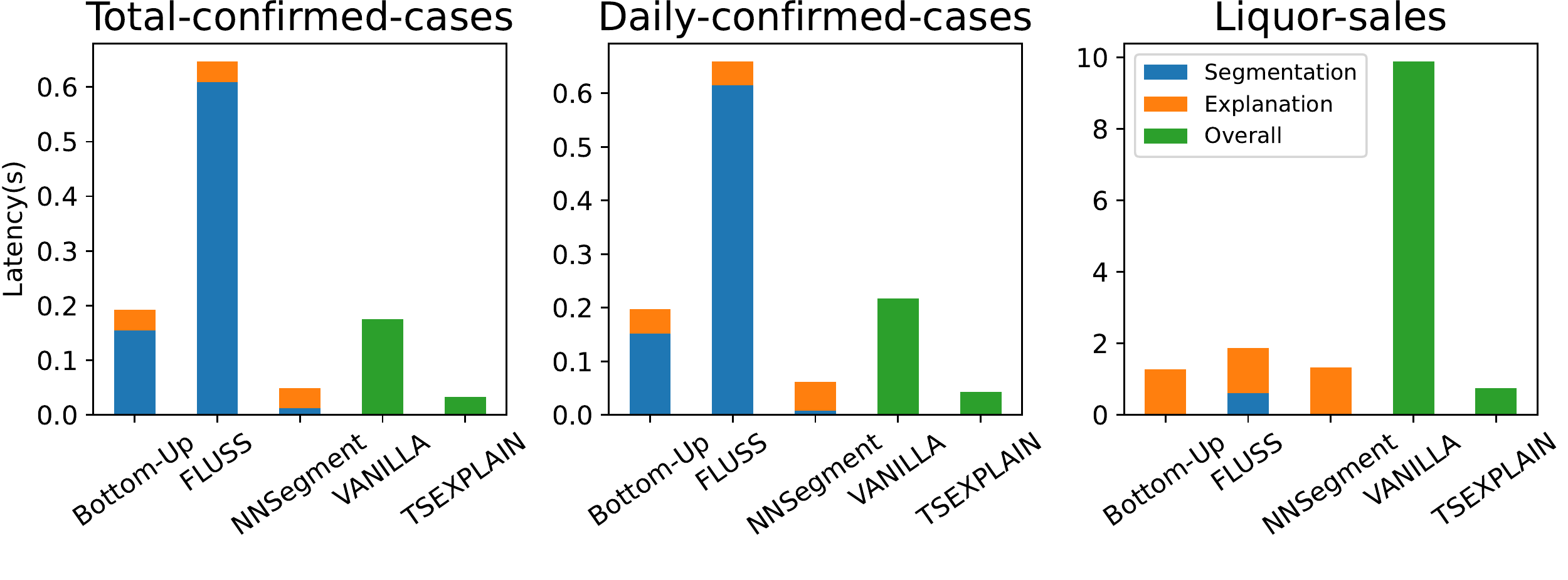}
\caption{Efficiency comparison with baselines.}
\label{fig:efficiency}
\end{figure}

\paragraph{{\textbf{Methods}}}
The three baselines in \Cref{sec:baseline} solely focus on visual shape based segmentation without providing any explanations and require the segmentation number as an input. To make them comparable, first, after segmenting using each baseline, we add the explanation module using the CA algorithm in \Cref{ssec:pipeline}; second, we reuse the optimal $K$ \sys finds in \Cref{ssec:expreal}, and then run all baselines and \sys with this given optimal $K$. We also remark that the latency of determining the optimal K is very low, around 2ms in our experiments.
\paragraph{{\textbf{Results}}}
\Cref{fig:efficiency} reports the end-to-end efficiency comparison. For each baseline, we report the segmentation and explanation time separately, while for \sys, we show the overall time since our segmentation module interleaves with the explanation module. To illustrate the effectiveness of our proposed optimizations, we also report {\vanSys} (VANILLA). We can tell that for different datasets, {\it FLUSS} is always the slowest, {\it NNSegment} and  {\it Bottom-Up} rank in the middle. {\vanSys} is similar to the {\it Bottom-Up} on the {\it COVID-19} datasets and becomes slow when the predicate number goes up in the {\it Liquor-sales} dataset.  Yet, combined with all the proposed optimizations, {\sys} is the fastest compared to all baselines on all datasets.

\subsubsection{Scalability}\quad \newline

\begin{figure}
    \center
    \includegraphics[width=.7\columnwidth]{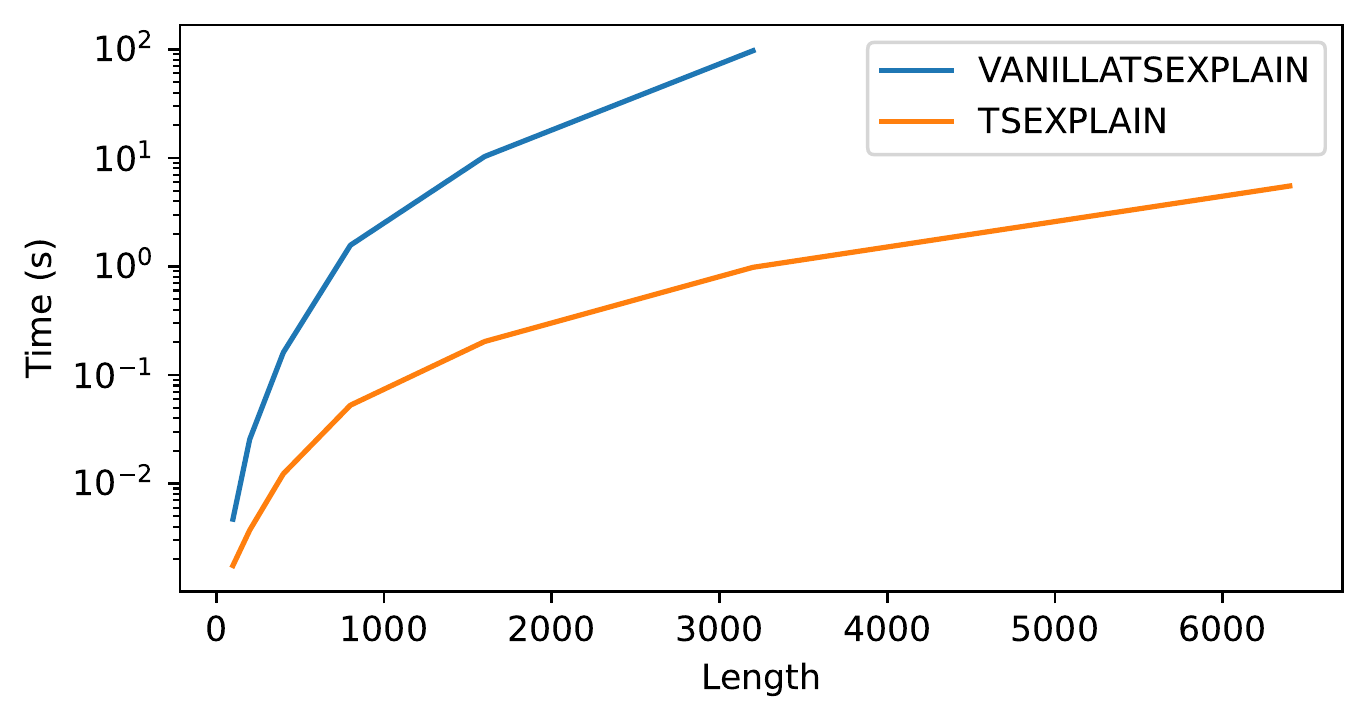}
    \caption{Different latency for time series of different lengths.}
    \label{fig: scale}
\end{figure}

 In the scalability experiment, we synthesized new time series following the procedure in Sec VII.A of different lengths = 100, 200, 400, 800, 1600, 3200, 6400. For each length, we synthesize five different time series.  We explain these time series using \vanSys (without any optimizations) and \sys with all optimizations mentioned in the paper.  \Cref{fig: scale} reports the average latency of different time series lengths. We terminate when the latency is greater than 100s. \vanSys's latency increases exponentially. With optimizations, \sys's latency increases much slower when the time series becomes longer. In particular, \sys can interactively explain the synthetic time series with length = 3200 in 982 ms. 

\stitle{Takeaway.}
\sys can explain these three time series within 800ms, and our optimizations have accelerated the running time up to 13$\times$ with neglectable effects on quality. What's more, \sys is faster than all the baselines.

%% file: content/discussion.tex
\section{Discussion}

\stitle{Time-varying Attribute.}
Different from non-temporal attributes, time-varying attributes are those whose values may change over time~\cite{jensen1996semantics}. Augmenting dataset with time-varying attributes can sometimes add insights in explaining trends. Below we demonstrate an example of using \sys with time-varying attributes.

\begin{figure}
    \centering
    \begin{subfigure}{\columnwidth}
        \centering
        \includegraphics[width=\columnwidth]{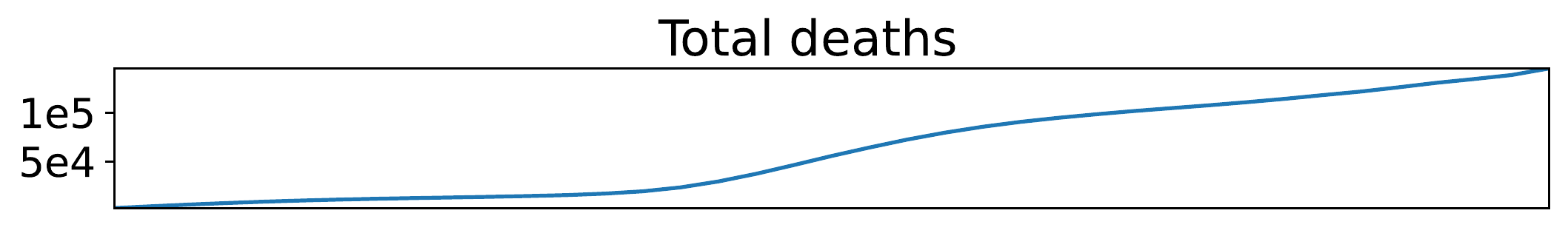}
    \end{subfigure}
    \begin{subfigure}{\columnwidth}
        \centering
        \includegraphics[width=\columnwidth]{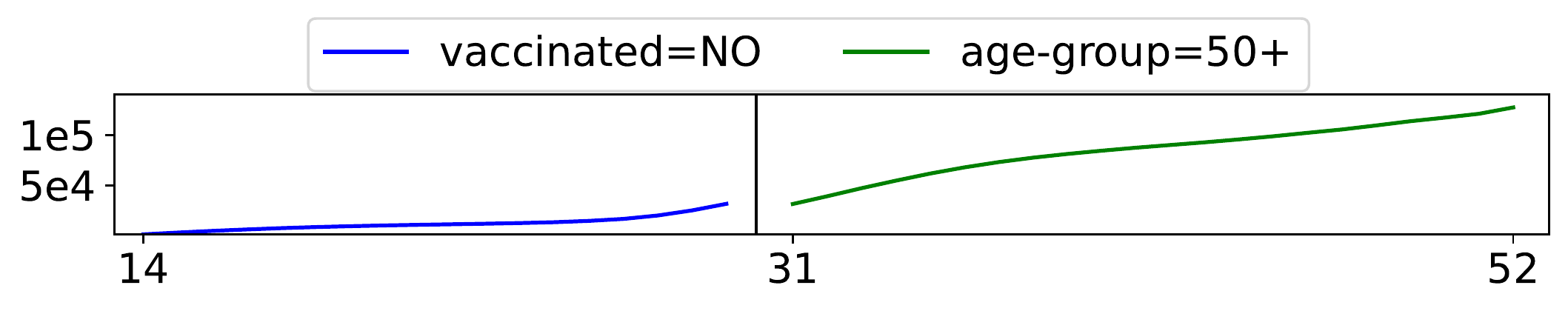}
    \end{subfigure}
    \caption{Segmentation of weekly total deaths. Overall trend(top). \sys results(bottom).}
    \label{fig:agegroup-death}
\end{figure}

The covid death dataset ~\cite{covid-deaths} records weekly deaths of different age groups and different vaccinated status from week 14 to 52 in 2021.  The {\tt vaccinated} attribute is a time-varying attribute since a person of {\tt vaccinated=NO} can shift to {\tt vaccinated=YES} in the near future. However, {\tt age-group} is a static attribute because one person belonging to a specific age group will not change within a year.  We explain the total death trend in \Cref{fig:agegroup-death}(top) using the attributes {\tt age-group} and {\tt vaccinated}. \Cref{fig:agegroup-death}(bottom) shows the result that before week 31, the main contributor is the unvaccinated people and after week 32, the main contributor shifts to elder people with {\tt age-group=50+}.  Combined with human knowledge that the vaccinated population is gradually becoming large along the time, we can get the insight that at the beginning, unvaccinated people (including unvaccinated elders) are the major factor of total deaths since unvaccinated young people also face high risk. Later on, elder people no matter vaccinated or not are the major reason as more young people get protected from vaccines while elder people do not get protected that well even vaccinated.

\stitle{Real-time Time Series.}
We briefly discuss how \sys can be extended to support real-time time series explanation. \sys first gives users the segmentation results of existing time series and meanwhile, caches all unit segments' top explanations. When new data arrives, it incrementally computes the top explanations for the new time series, runs the segmentation algorithm based on the existing time series' cutting point and newly arrived data points, and updates the segmentation results.

\stitle{Seasonal Datasets}
For seasonality datasets, \sys can explain the seasonality dataset directly and detect the repeated pattern of evolving explanations which indicates the periodicity property. Users can also first decompose the seasonal datasets~\cite{hyndman2018forecasting} and explain the seasonality and trend separately.

%% file: content/conclusion.tex
\section{Conclusion}
This work introduces \sys, the first explanation engine that identifies the evolving explanations for aggregated time series.
We formulate the problem for deriving evolving explanations as a {\it K-Segmentation} problem, aiming to partition the input time series into K smaller segments such that each period has consistent top explanations. We propose a novel variance metric to quantify the consistency in each segment and develop a dynamic programming algorithm for identifying the optimal {\it K-Segmentation} scheme. \sys can automatically identify the optimal $K$ using the "elbow method". In the experiments, we show \sys can effectively discover the evolving explanation on synthetic and real-world datasets. We propose optimizations that enable \sys to answer all our queries interactively within one second.
Several future work directions include extending the difference metric library, recommending explain-by attributes, adding hints for segments with higher variance for further inspection.